\documentclass[12pt,a4paper]{article}

\usepackage{epsfig}
\newcommand{\bi}{\begin{itemize}}
\newcommand{\ei}{\end{itemize}}
\newcommand{\be}{\begin{equation}}
\newcommand{\ee}{\end{equation}}

\newcommand{\bea}{\begin{eqnarray}}
\newcommand{\eea}{\end{eqnarray}}
\newcommand{\beastar}{\begin{eqnarray*}}
\newcommand{\eeastar}{\end{eqnarray*}}

\newcommand{\eq}[1]{(\ref{#1})}
\newcommand{\eqq}[2]{(\ref{#1},\ref{#2})}
\newcommand{\eqqq}[3]{(\ref{#1},\ref{#2},\ref{#3})}

\newcommand{\s}{\sigma}
\newcommand{\eps}{\epsilon}

\begin{document}

\title{Work and heat fluctuations in two-state systems: a trajectory
thermodynamics formalism}

\author{F. Ritort\\{\em\small Departament de Fisica Fonamental,
Facultat de F\'{\i}sica, Universitat de Barcelona,}\\ {\em\small Diagonal
647, 08028 Barcelona, Spain}\\{\small\tt E-Mail:ritort@ffn.ub.es, http://www.ffn.ub.es/ritort}}

\maketitle

\abstract{ Two-state models provide phenomenological descriptions of
many different systems, ranging from physics to chemistry and
biology. We investigate work fluctuations in an ensemble of two-state
systems driven out of equilibrium under the action of an external
perturbation. We calculate the probability density $P_N(W)$ that a work
equal to $W$ is exerted upon the system (of size $N$) along a given
non-equilibrium trajectory and introduce a trajectory thermodynamics
formalism to quantify work fluctuations in the large-$N$ limit. We then
define a {\em trajectory entropy} $S_N(W)$ that counts the number of
non-equilibrium trajectories $P_N(W)=\exp(S_N(W)/k_BT)$ with work equal
to $W$ and characterizes fluctuations of work trajectories around the
most probable value $W^{\rm mp}$. A {\em trajectory free-energy} ${\cal
F}_N(W)$ can also be defined, which has a minimum at $W=W^{\dag}$, this
being the value of the work that has to be efficiently sampled to
quantitatively test the Jarzynski equality. Within this formalism a
Lagrange multiplier is also introduced, the inverse of which plays the
role of a {\em trajectory temperature}.  Our general solution for
$P_N(W)$ exactly satisfies the fluctuation theorem by Crooks and allows
us to investigate heat-fluctuations for a protocol that is invariant
under time reversal. The heat distribution is then characterized by a
Gaussian component (describing small and frequent heat exchange events)
and exponential tails (describing the statistics of large deviations and
rare events). For the latter, the width of the exponential tails is
related to the aforementioned {\em trajectory temperature}. Finite-size
effects to the large-$N$ theory and the recovery of work distributions
for finite $N$ are also discussed.  Finally, we pay particular attention
to the case of magnetic nanoparticle systems under the action of a
magnetic field $H$ where work and heat fluctuations are predicted to be
observable in ramping experiments in micro-SQUIDs.}

\newpage

\baselineskip 15pt

\section{Introduction}
\label{intro}
There has been recent interest in the experimental measure of work
fluctuations and the test of the so-called fluctuation
theorems. Initially proposed in the context of sheared systems in a
steady state~\cite{EvaCohMor93} several versions of such theorems have
been derived~\cite{EvaSea02}. In particular, specific identities have
been obtained in the context of stochastic systems that show how it is
possible to recover the equilibrium free-energy change in a reversible
transformation by exponential averaging over many non-equilibrium
trajectories that start at
equilibrium~\cite{Jarzynski97,Kurchan98,Crooks98}.  Let us consider a
system initially in equilibrium in contact with a thermal bath (at
temperature $T$) that is submitted to an isothermal perturbation
according to a given protocol. Work fluctuations (WF) refer to the fact
that the work $W$ exerted upon the system depends on the particular
non-equilibrium trajectory followed by the system. As the initial
configuration or the trajectory are stochastic, the value of the work
$W$ changes among different trajectories, all generated with the same
perturbation protocol.  Transient violations (TV) of the second law
refer to the fact that, among all possible WF, a fraction of them absorb
heat from the bath that is transformed into work. Taken individually,
these rare trajectories violate the Clausius inequality, $Q\le T\Delta
S$, where $Q$ is the heat supplied from the bath to the system and
$\Delta S$ is the change in the entropy, a state function defined
through the transformation. In a transformation cycle $\Delta S=0$ these
TV satisfy $Q>0$ i.e., they can absorb a net amount of heat from the
bath during the cycle.  In terms of the dissipated work $W_{\rm dis}$,
the Clausius relation can be expressed in the following form,
\be
W_{\rm dis}=W-W_{\rm rev}=T\Delta S- Q\ge 0
\label{intro1}
\ee
In this expression $W$ is the total work exerted upon the
system. According to the first law of thermodynamics (conservation of
the energy) $W$ is given by $W=\Delta E-Q$ where $\Delta E$ is the
change in the internal energy, $W_{\rm rev}$ is the reversible work
(identical to the free-energy change $\Delta F=\Delta E-T\Delta
S$). Both the heat $Q$ and $W_{\rm dis}$ (or $W$) are trajectory
dependent, however $\Delta S$ and $W_{\rm rev}$ are both trajectory
independent as they are state functions, only dependent on the initial
and final states. The Clausius inequality \eq{intro1} has to be
understood as a result that is valid after averaging the fluctuating
quantities $Q$ and $W$ over an infinite number of trajectories (in what
follows we will denote this average by $\overline{(..)}$).  The second
law reads $\overline{W_{\rm dis}}\ge 0$ and TV of the second law refer
to the existence of trajectories where $W_{\rm dis}<0$. From this point
of view, TV are just WF characterized by the fact that $W_{\rm
dis}<0$. The interest in studying TV is that these describe large
deviations of the work that have to be sampled in order to recover
equilibrium free-energy differences from non-equilibrium
measurements~\cite{HumSza01}. 

The steadily increasing development of nanotechnologies during the last
decade has made WF experimentally accessible. Recent experiments on
single RNA hairpins unfolded under the action of mechanical
force~\cite{LipDumSmiTinBus02} and micro-sized beads trapped by laser
tweezers and moved through a solvent~\cite{WanSevMitSeaEva02} have
provided a first quantitative estimate of WF and TV. Related
measurements include the experimental verification of the
Gallavotti-Cohen fluctuation theorem in Rayleigh-Bernard convection
\cite{CilLar98} and turbulent flows \cite{CilGarHerLacPinRui03}.  This
research is potentially very interesting as it leads to new insights
about the physical processes occurring at the nanoscale, a frontier that
marks the onset of complex organization of
matter~\cite{LauPinSchStoWol00}. A characteristic of WF is that they are
quickly suppressed as the system size or the time window
of the measurement increase.

The central quantity describing WF is the work probability
distribution $P_N(W)$ ($N$ stands for the system size), $P_N(W)dW$
being the fraction of non-equilibrium trajectories with work between
$W$ and $W+dW$. The knowledge of this quantity is important for what
it tells us about the mathematical form of the tails of the
distribution, relevant to understand the importance of large
deviations of work values respect to the average value. A precise
knowledge of the form of the tails in that distribution gives us hints
about how many experiments need to be done in order to recover
equilibrium quantities from non-equilibrium experiments.  In this work
we investigate an ensemble of two-level systems as an explicit example
where $P_N(W)$ can be analytically computed in the large-$N$ approach
using a path integral method. This approach allows us to exactly
derive several exact results describing work and heat fluctuations in
the system in the large-$N$ limit but also for finite $N$. The most
important result in the paper is the introduction of a trajectory
thermodynamics formalism, the key quantity being the {\em trajectory
entropy} $S_N(W)$. This allows us to infer several quantities such as
the {\em trajectory free-energy} ${\cal F}_N(W)$ and the {\em
trajectory temperature} $\lambda(W)$, the latter being a Lagrange
multiplier that plays the role of the inverse of a temperature, an
intensive variable related to the statistics of large deviations or
tails in the work and heat distributions. Two-state models represent a
broad category of systems where WF and TV can be predicted to be
experimentally observable making the present calculations relevant as
they might allow a detailed comparison between theory and
experiments. In particular, we propose magnetic nanoparticles as
excellent candidate systems to experimentally test the present theory.

The plan of the paper is as follows. In Sec.~\ref{model} we describe the
model and the large-$N$ approach. In Sec.~\ref{thermo} we develop the
trajectory thermodynamics formalism that allows us to reconstruct the
work distribution, define a {\em trajectory entropy}
$S_N(W)=\log(P_N(W))$ and a {\em trajectory free-energy} ${\cal
F}_N(W)$. In Sec.~\ref{numerics} we show how the saddle-point equations
derived in Sec.~\ref{model} can be numerically solved. The dependence of
the main parameters of the theory (most probable work $W^{\rm mp}$,
transient violations work $W^{\dag}$ and fluctuation-dissipation ratio
$R$) on the field protocol are discussed in Sec.~\ref{mp}.  Within the
formalism it is then possible to show, Sec.~\ref{ft}, that the entropy
per particle $s(w)$ ($w$ being the work $W$ per particle) exactly
satisfies the fluctuation theorem by Crooks. Moreover, it is possible to
infer the shape of the tails in the work distribution from the sole
knowledge of the Lagrange multiplier conjugated to the {\em trajectory
entropy}, $\lambda(w)$, that plays the role of the inverse of a
temperature (what we call the {\em trajectory temperature}) in the
formalism. In Sec.~\ref{heat} we study heat fluctuations in the
model. We show the existence of two sectors in the heat distribution
that are described by a Gaussian central part (corresponding to small
and most probable deviations) and two exponential tails (corresponding
to large and rare deviations) showing the presence of intermittent heat
fluctuations in the theory. In Sec.~\ref{fse} we discuss finite-size
corrections to the large-$N$ theory and how $P_N(W)$ for finite $N$ can
be reconstructed using the results from the large-$N$ approach.
Particular emphasis is finally placed in Sec.~\ref{nano} in the case of
magnetic nanoparticle systems where WF are predicted to be
experimentally observable and described by the present theory.
Sec.~\ref{conclusions} presents the conclusions.

\section{Ensemble of two-state
systems: the large-$N$ approach}
\label{model}
A broad category of systems can be modeled by an ensemble
or collection of independent two-state systems. These offer realistic
descriptions of electronic and optical devices that can function in
two different configurations, atoms in their ground and excited
states, magnetic particles whose magnetic moment can point in two
directions, or biomolecules in their native and unfolded states, among
others. Throughout the paper, and in view of the possible experimental
implications, we will adopt the nomenclature of magnetic systems.  A
particle $i$ in the ensemble ($1\le i\le N$) has magnetic moment $\mu$
and can point in two directions according to the sign of the spin
$\s_i=\pm 1$. A given configuration in the ensemble is specified by a
string of spin values ${\cal
C}\equiv\lbrace\s_1,\s_2,..,\s_N\rbrace$. In the presence of an
external field $H$, the energy of a configuration ${\cal C}$ is given
by
\be
E({\cal C})=-\mu H M({\cal C})= -\mu H\sum_{i=1}^N\s_i~~~~~,
\label{a1}
\ee
$M({\cal C})=\sum_{i=1}^N\s_i$ being the total magnetization of the
system. The transition rates for individual particles will
be denoted as $p^{\rm up}(H),p^{\rm down}(H)$ to indicate the
transitions $\s=-1\to\s'=1$ and $\s=1\to\s'=-1$ respectively.  These
rates satisfy detailed balance, therefore $p^{\rm up}(H)/p^{\rm
down}(H)=\exp(-2\beta \mu H)$ where $\beta=1/k_BT$, $T$ being the bath
temperature and $k_B$ the Boltzmann constant. The overall transition
rate is given by $p^{\rm tot}(H)=p^{\rm up}(H)+p^{\rm
down}(H)$. Although it is possible to introduce structural disorder in
the ensemble (e.g. by allowing $\mu$ or $p^{\rm up}(H)$ to be a random
quenched variable), in the following analysis we will restrict us to
the non-disordered or mono-disperse case.

Let the system be prepared at $t=0$ in an equilibrium state at an
initial value of the field $H_0=H_i$ and let us consider an external
isothermal perturbation that changes the field $H$ according to a
protocol function $H(t)$. Throughout this paper we will denote this
non-equilibrium process a {\em ramping} experiment. If the variation
is slow enough then the process is quasi-static and the system goes
through a sequence of equilibrium states. However, if the rate
$\dot{H}$ is large compared to the relaxation time of the particle
then the magnetization $M=\sum_{i=1}^N\s_i$ does not follow the
equilibrium curve $M_{\rm eq}(H)=N\tanh(\beta\mu H)$. To specify a
trajectory it is then convenient to discretize time in $N_s$
time-steps of duration $\Delta t$ each and take the continuous-time
limit $\Delta t \to 0,N_s\to\infty$ (with the total time $t=N_s\Delta
t$ fixed) at the end. The perturbation protocol is specified by the
sequence of values $\lbrace H_k; 1\le k\le N_s\rbrace$ and a
trajectory ${\cal T}$ is defined by the sequence of configurations
${\cal T}=({\cal C}_k;1\le k\le N_s)$ where ${\cal
C}_k=\lbrace\s_i^k;1\le i\le N\rbrace$ is the configuration at time
$t=k\Delta t$. The total work exerted upon the system along a given
trajectory is given by~\cite{Crooks98},
\be
W({\cal T})=-\mu\sum_{k=0}^{N_s-1} M_{k+1} (H_{k+1}-H_k)
\label{a2}
\ee
$M_k=\sum_{i=1}^N\s_i^k$ being the magnetization at time-step $k$.  The
dissipated work for a given trajectory is the difference between the
total work and the reversible one, $W_{\rm dis}=W-W_{\rm rev}$ where
$W_{\rm rev}=\Delta F$ is the change in equilibrium free-energy between
the initial and final values of the field. The free energy is given by
$F(H)=-Nk_B T \log(2\cosh(\beta\mu H))$. To quantify WF we have to
compute the probability distribution for the total work measured over
all possible non-equilibrium trajectories,
\be
P_N(W)=\sum_{{\cal T }}p({\cal T})\delta(W-W({\cal T}))=\sum_{\lbrace
\s_i^k\rbrace} p({\cal T})\delta(W+\mu\sum_{k=0}^{N_s-1} M_{k+1} (H_{k+1}-H_k))~,
\label{a3}
\ee
where $p({\cal T})$ denotes the probability of a given trajectory. The
subindex $N$ in $P_N(W)$ is written to emphasize the dependence of the
distribution on the size of the system. $P_N(W)$ is computed using the
Bayes formula $p({\cal T})=\prod_{k=0}^{N_s-1}q_k(\lbrace\s_i^{k+1}
\rbrace| \lbrace\s_i^{k} \rbrace) p_0(\lbrace\s_i^{0}\rbrace)$, where
$q_k(\lbrace\s' \rbrace| \lbrace\s\rbrace)$ denotes the transition
probability to go from $\lbrace\s\rbrace$ to $\lbrace\s'\rbrace$ at
time-step $k$, and $p_0(\lbrace\s_i^{0}\rbrace)$ is the initially
equilibrated (i.e. Boltzmann-Gibbs) distribution. Evaluation of the
integral \eq{a3} requires the following steps: 1) trace out spins in the
sum; 2) insert the factorized expression for $p({\cal T})$; 3) use the
integral representation for the delta function
$\delta(x)=(1/2\pi)\int_{-\infty}^{\infty}dx\exp(i\lambda x)$ and 4)
insert the following factor,
\be
1=\prod_{k=0}^{N_s-1}\frac{1}{2\pi}\int_{-\infty}^{\infty} d\gamma_k dM_k\exp(i\gamma_k(M_k-\sum_{i=1}^N\s_i^k))~~~.
\label{a4}
\ee
After some manipulations this leads to the following expression for the
work probability distribution (up to some unimportant multiplicative terms),
\be
P_N(W)\propto \int d\lambda \prod_{k=0}^{N_s-1}(d\gamma_k dm_k)\exp\Bigl(
A(w,\lambda,\lbrace\gamma_k\rbrace,\lbrace m_k
\rbrace)\Bigr)
\label{a5}
\ee
where $A$ is the saddle-point function, $w=W/N,m_k=M_k/N$ (throughout the
paper we will use small case letters to refer to intensive quantities).  
The function $a=A/N$ is given by,
\bea
a(w,\lambda,\lbrace\gamma_k\rbrace,\lbrace m_k \rbrace)=-\lambda
\Bigl(w+\mu\sum_{k=0}^{N_s-1}m_{k+1}(H_{k+1}-H_k)\Bigr)-\sum_{k=0}^{N_s}\gamma_k
m_k+\nonumber\\
\sum_{k=0}^{N_s-1}\Bigl(
\frac{m_k+1}{2}\log(u_{k+1})+ \frac{1-m_k}{2}\log(v_{k+1}) \Bigr)+\log(e^{\gamma_0}p^{\rm up}(H_i)+e^{-\gamma_0}p^{\rm down}(H_i))~~~.
\label{a6}
\eea
The terms $u_k,v_k$ are given by,
\bea
u_{k+1}=\exp(\gamma_{k+1})(1-p_k^{\rm down})+\exp(-\gamma_{k+1})p_k^{\rm down}\\
v_{k+1}=\exp(\gamma_{k+1})p_k^{\rm up}+\exp(-\gamma_{k+1})(1-p_k^{\rm up}).
\eea
with the boundary condition $\gamma_{N_s}=0$. The quantities $p^{\rm
up}(H_i),p^{\rm up}(H_i)$ are the transition rates at time $s=0$, and we
are assuming that at the initial condition the system is in thermal
equilibrium.
In the continuous-time limit \eq{a5} 
becomes a path integral over the variable $\lambda$ and the functions
$\gamma(t),m(t)$ with,
\bea 
a(w,\lambda,\gamma(s),m(s))=-\lambda
\Bigl(w+\mu\int_0^tm(s)\dot{H}(s)ds\Bigr)
+\nonumber\\\frac{1}{2}\int_0^t \Bigl(m(s)(2\dot{\gamma}(s)+c(s))+d(s)\Bigr)ds
+\log(e^{\gamma(0)}p^{\rm up}(H_i)+e^{-\gamma(0)}p^{\rm down}(H_i))
\label{a9}
\eea
where 
\bea
c(s)=p^{\rm down}(s)(\exp(-2\gamma(s))-1)-p^{\rm
up}(s)(\exp(2\gamma(s))-1) \label{a10a}\\
d(s)=p^{\rm down}(s)(\exp(-2\gamma(s))-1)+p^{\rm up}(s)(\exp(2\gamma(s))-1)~~~~.\label{a10b}
\eea
As we are interested in the crossover to the large-$N$ regime we can
estimate the integral \eq{a5} by using the saddle-point method. For each
value of the work trajectory $w$ the dominant contribution is given by
the solution of the functional equations,
\bea
\frac{\delta a}{\delta \lambda}=w+\mu\int_0^tm(s)\dot{H}(s)ds=0\label{a11a}\\
\frac{\delta a}{\delta \gamma(s)}=\dot{m}(s)+m(s)p^{\rm tot}(s)-(p^{\rm
up}(s)-p^{\rm down}(s))+m(s)d(s)+c(s)=0\label{a11b}\\
\frac{\delta a}{\delta m(s)}=\dot{\gamma}(s)-\lambda\mu \dot{H}(s)+\frac{1}{2}c(s)=0\label{a11c}
\eea
with the boundary conditions 
\be
\gamma(t)=0~~~;~~~m(0)=\tanh(\gamma(0)+\beta\mu H_i)~~~. 
\label{bound}
\ee
Note that the boundary conditions are a bit special as causality is
broken. The function $\gamma(s)$ has the boundary condition located at
the final time $s=t$ while the boundary condition for $m(s)$ is located
at the initial time $s=0$.  These equations can be numerically solved in
general and analytically solved only partially and for some particular
cases (e.g. in the case where the rate $\dot{H}$ is constant). Before
presenting detailed numerical solutions to these equations we should
point out several general aspects of such solutions.  At first we note
how, for a given value of $\lambda$, Eq.~\eq{a11c} together with the
boundary condition $\gamma(t)=0$ can be solved giving the solution
$\gamma_{\lambda}(s)$, the subindex $\lambda$ emphasizing the dependence
of this solution on the parameter $\lambda$. Inserting this result in
\eq{a11b} and using the boundary condition \eq{bound} we get the
solution $m_{\lambda}(s)$. Finally, insertion of $m_{\lambda}(s)$ in
\eq{a11a} gives a value for the work $w(\lambda)$. This last relation
can then be inverted to give $\lambda(w)$ and from it, the solutions
$\gamma_{\lambda}(s),m_{\lambda}(s)$ will also depend on the value of
$w$. To better emphasize this dependence we will denote by
$\lambda(w),\gamma_w(s),m_w(s)$ the solutions of \eqqq{a11a}{a11b}{a11c}
for a given value of $w$ and
\be
s(w)=a(w,\lambda(w),\gamma_w(s),m_w(s))
\label{snold}
\ee
the corresponding extremal value of $a$. We will also make explicit the $w$-dependence in the
time-dependent quantities $c(s),d(s)$ in \eqq{a10a}{a10b} and denote
them by $c_w(s),d_w(s)$ respectively.  Furthermore, we can define the
trajectory entropy $S_N(W)$,
\be
P_N(W)=\exp(S_N(W))~~~~.
\label{snw}
\ee
In the large-$N$ limit, from \eqq{a5}{snold} we have
\be
s(w)=\lim_{N\to\infty}\frac{S_N(W)}{N}~~~{\rm with}~~~ W=Nw~~~,
\label{snw2}
\ee 
the function $s(w)$ playing the role of a trajectory entropy per
particle that counts the density of trajectories per particle with work
equal to $w$.  This means that, for $N$ finite,
$\Phi_N(w)dw=\exp(Ns(w))dw$ is approximately proportional to the
fraction of trajectories with work between $w$ and $w+dw$. From
\eqq{snw}{snw2} an approximate expression for the work probability
distribution can be written,
\be P_N(w)=\frac{\Phi_N(w)}{\int_{w_{\rm min}}^{w_{\rm
max}}\Phi_N(w')dw'}=\frac{\exp(Ns(w))}{\int_{w_{\rm min}}^{w_{\rm
max}}\exp(Ns(w'))dw'}
\label{distrib}
\ee
where $w_{\rm min}$ and $w_{\rm max}$ are the minimum and maximum
possible values of the work. Clearly, from \eq{a2} these values are
given by $w_{\rm max}=-w_{\rm min}=\mu (H_f-H_i)$ where $H_f=H(t)$ is
the final value of the magnetic field. The subindex $N$ in $P_N(w)$ and
$\Phi_N(w)$ emphasize the dependence of these quantities on the size of
the system. Finally, we note that, albeit the solutions
\eqqq{a11a}{a11b}{a11c} have been obtained using the saddle-point
approximation (only valid for large $N$) the final result \eq{distrib}
can be very accurate for small values of $N$. This result, that at
first glance may appear striking, is just consequence of the non-interacting
character of the Hamiltonian \eq{a1}. This point is discussed in more
detail in Sec.~\ref{fse}. There we show that, albeit \eq{distrib} is
only approximate for finite $N$, the cumulants that we can extract from
$s(w)$ are exact for any $N$. This allows us to exactly reconstruct
the finite $N$ distribution from the sole knowledge of $s(w)$.

The action $A=Na$ in \eq{a9} could be used (employing Monte Carlo
algorithms) to generate trajectories according to their probability
$P_N(w)$~\footnote{The easiest procedure then would be to start from an
initial trajectory $\gamma(s),m(s)$ (satisfying the boundary conditions
$m(0)=\tanh(\gamma(0)+\beta\mu H_i);\gamma(t)=0$) and perform successive
``local'' updates along the trajectory and accepting the moves according
to the change in the action $A$ (by using an algorithm that satisfies
detailed balance, as defined by the action $A$, and respects the boundary
conditions).}. Inserting \eq{a11c} in \eq{a9} we get,
\be
s(w)=-\lambda w+\frac{1}{2}\int_0^t d_w(s)ds~~~~.
\label{as}
\ee
The value $w$ for which $s(w)$ is maximum yields the most probable work
($w=w^{\rm mp}$) among all trajectories. This can be evaluated using the
equation
\be
s'(w)=\frac{\partial a}{\partial w}=-\lambda(w)~~~~,
\label{maximum} 
\ee
where we have used the chain rule together with the extremum conditions
\eqqq{a11a}{a11b}{a11c} as well as \eq{a9}. We will see later in
Sec.~\ref{heat} that the Lagrange multiplier $\lambda(w)$ is related to
the inverse of a new energy scale or temperature that describes the
tails of the work distribution. This quantity is of much current
interest as it describes the statistics of rare events and large
deviations of work values from the average which are observable in small
systems. The extremum solution of \eq{maximum} can then be written as
$\lambda(w^{\rm mp})=0$,
\be
\frac{\partial s(w)}{\partial w}\Bigr|_{w=w^{\rm mp}}=0 ~~~~~{\rm or}~~~~~
\lambda(w^{\rm mp})=0~~~.
\label{maximum2}
\ee
This solution solves \eqqq{a11a}{a11b}{a11c} giving $\gamma_{w_{\rm
mp}}(s)=c_{w^{\rm mp}}(s)=d_{w^{\rm mp}}(s)=0$. Eqs.\eqq{a11a}{a11b} then
give the solution for the most probable trajectory (usually derived
using standard statistical methods),
\be
\dot{m}(s)=-m(s)p^{\rm tot}(s)+(p^{\rm up}(s)-p^{\rm down}(s))~~~~.
\label{nn1}
\ee
The reversible process is a
special case (only valid for slow enough perturbation protocols) and
corresponds to $\dot{m}(s)=0$ or 
\be
m(s)=(p^{\rm up}(s)-p^{\rm down}(s))/p^{\rm tot}(s)=\tanh (\beta \mu H(s))~~~~.
\label{nn2}
\ee

\section{Trajectory thermodynamics formalism} 
\label{thermo}
From the trajectory entropy $s(w)$ we can construct a trajectory free-energy ${\cal F}(w)$ useful to predict under which conditions TV are
properly sampled and fluctuation theorems can be quantitatively verified. For this we
consider the Jarzynski equality \cite{Jarzynski97},
\be
\overline{\exp\Bigl(-\frac{W}{k_BT}\Bigr)}=\exp\Bigl(-\frac{\Delta F}{k_BT}\Bigr)
\label{0}
\ee
that we can write as,
\bea
\exp\Bigl( -\frac{\Delta F}{k_BT}\Bigr)=\int dW P_N(W)\exp\Bigl(-\frac{W}{k_BT} \Bigr)=\nonumber\\
\int dW \exp\Bigl(-\frac{W}{k_BT} +S_N(W)\Bigr)=\int dW
\exp\Bigl(-\frac{{\cal F}_N(W)}{k_BT}\Bigr)
\label{00}
\eea
where we used \eq{snw} and we have defined the trajectory free-energy,
\be
{\cal F}_N(W)=W-k_BTS_N(W)~~~.
\label{11}
\ee
In the large-$N$ limit, using \eq{snw2}, we can write
\bea
\exp\Bigl( -\frac{\Delta F}{k_BT}\Bigr)=\int dw
\exp\Bigl(-\frac{N}{k_BT}(w-k_BT s(w))\Bigr)=\nonumber\\
\int dw\exp\Bigl(-\frac{N{\cal F}(w)}{k_BT}\Bigr)
\equiv
\exp\Bigl(-\frac{N{\cal F}(w^{\dag})}{k_BT}\Bigr)
\label{eqfree}
\eea
where 
\be
{\cal F}(w)=w-k_BT s(w)
\label{eqfree2}
\ee
is a trajectory free-energy (per particle) that depends on the
particular value of the work $w$. Evaluating the integral \eq{eqfree} by the
steepest descent method and using \eq{as} we obtain the {\em
thermodynamic} relations,
\bea
\frac{1}{k_BT}=\frac{\partial s(w)}{\partial w}\Bigr|_{w=w^{\dag}}=
-\lambda(w^{\dag})\label{eqmicro1}\\
{\cal F}(w^{\dag})=\Delta F/N=w_{\rm rev}=w^{\dag}-k_BTs(w^{\dag})=
-\frac{k_BT}{2}\int_0^t d_{w^{\dag}}(s)ds
\label{eqmicro2}
\eea
Using the definition \eq{eqfree2} together with \eqq{maximum2}{eqmicro1} we have the
relations,
\bea
\frac{\partial {\cal F}(w)}{\partial w}\Bigr|_{w=w^{\rm mp}}=1\label{rel1}\\
\frac{\partial {\cal F}(w)}{\partial w}\Bigr|_{w=w^{\dag}}=0\label{rel2}
\eea
i.e. the entropy has a maximum at $w=w^{\rm mp}$ and the free energy has
a minimum at $w=w^{\dag}$.  These relations bear similarity to those
considered in thermodynamics but now applied to work trajectory values. For
the case of the canonical ensemble the quantities $s(w),{\cal F}(w),w$
play the role of the standard entropy, free energy and internal energy
while $\lambda(w)$ is the intensive variable corresponding to the inverse
of a temperature.

\begin{figure}
\begin{center}
\epsfig{file=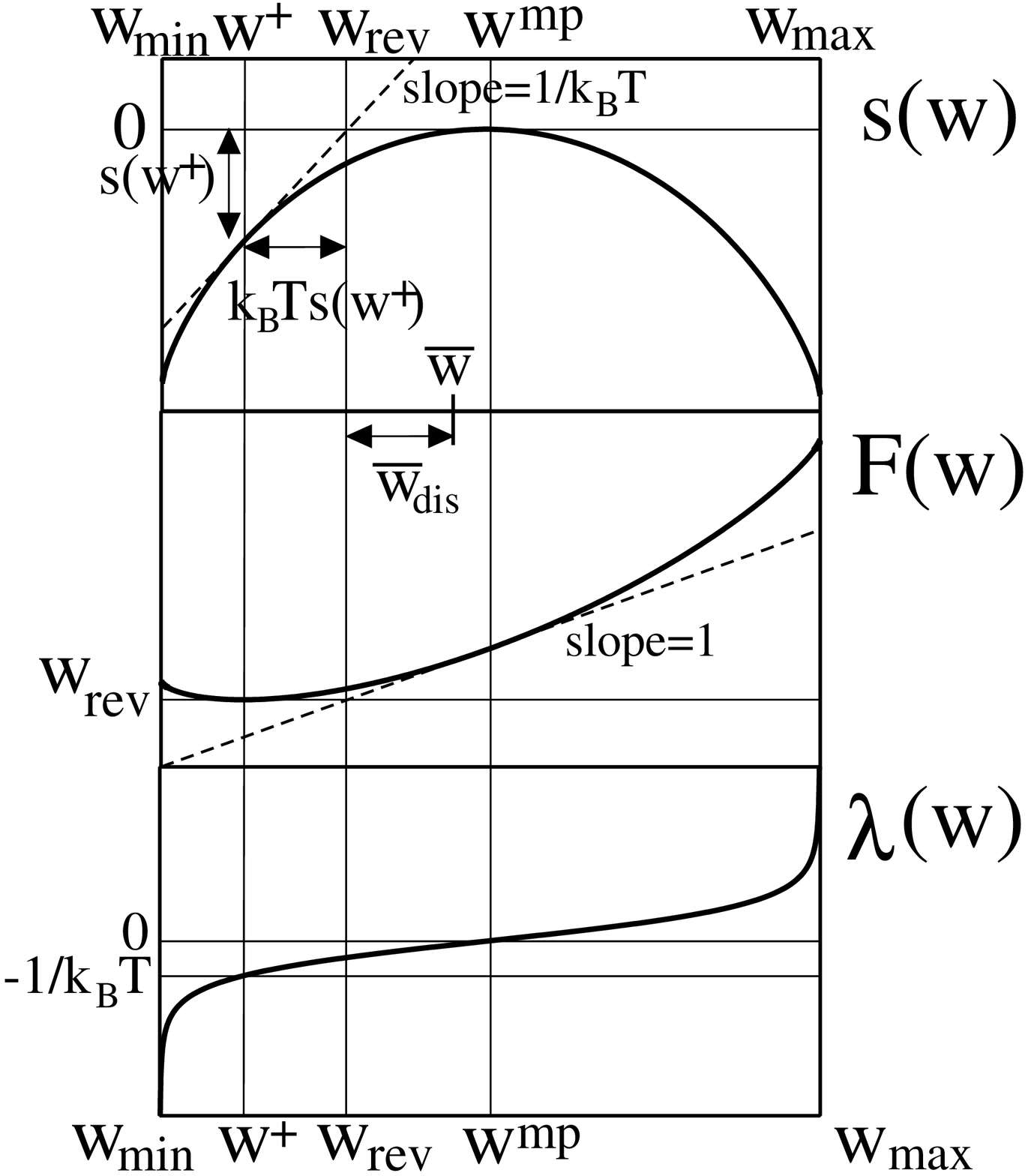,angle=-90, width=18cm}
\end{center}
\caption{\small Diagrams showing the different relevant quantities in the
trajectory thermodynamics formalism. Upper panel: Trajectory entropy
$s(w)$ related by \eq{distrib} to the density of trajectories with work
equal to $w$. Middle panel: Trajectory free-energy ${\cal
F}(w)=w-k_BTs(w)$. Lower panel: Lagrange multiplier $\lambda(w)$.  Six
are the most relevant work values: $w_{\rm max}$ and $w_{\rm min}$ for
the maximum and minimum values of the work; $w^{\rm mp}$ the most
probable work value given by $s'(w^{\rm mp})=\lambda(w^{\rm mp})=0$ or
${\cal F}'(w^{\rm mp})=1$; $w^{\dag}$ the value of the work that has to
be sampled to recover free energies from non-equilibrium work values
using the Jarzynski equality \eq{0}. This is given by
$s'(w^{\dag})=-\lambda(w^{\dag})=1/k_B T$ or ${\cal F}'(w^{\dag})=0$;
$w_{\rm rev}=F(H_f)-F(H_i)$ the reversible work; and $\overline{w_{\rm
dis}}=\int_{w_{\rm min}}^{w_{\rm max}} (w-w_{\rm rev})P_N(w)dw$ the
average dissipated work. They are related by $w_{\rm min}< w^{\dag}<
w^{\rm mp}<w_{\rm max}$ while the second law of thermodynamics imposes
$w_{\rm dis}\ge 0$.}
\label{fig1}
\end{figure}

A graphical construction of the relations
\eqq{eqmicro1}{eqmicro2} is shown in Figure~\ref{fig1}. This figure
illustrates how the most important quantities $w_{\rm rev},w^{\rm
mp},w^{\dag},\overline{w}$ are related to each other. In particular,
$\overline{w}=\int dw w P_N(w)$ is expected to differ from $w^{\rm mp}$
albeit that difference can be small for highly symmetric distributions.

The difference between $w^{\dag}$ and $w^{\rm mp}$ indicates that the
average \eq{eqfree} is properly weighed whenever trajectories with work
values around $w^{\dag}$ are sampled.  This result indicates that proper
sampling of non-equilibrium work values around $w^{\dag}$ is required to
derive equilibrium free-energies from non-equilibrium measurements by
using the Jarzynski equality.  A proper sampling of work values around
$w^{\dag}$ can be guaranteed when, out of the total number of
trajectories, a finite fraction of work trajectory values in the
vicinity of $w^{\dag}$ is observed. From a practical point of view this
means that the histogram of work values must extend down to $w^{\dag}$. If
this is not achieved, then the exponential average performed over a
finite number of non-equilibrium experiments has a bias that can
be estimated in some
cases~\cite{ZucWoo02,GorRitBus03}. Eqs.\eqq{eqmicro1}{eqmicro2} are
readily solved at the Gaussian level (i.e. assuming that $P_N(w)$ is
exactly a Gaussian or $s(w)$ a quadratic function) giving
$w^{\dag}=w^{\rm mp}-\s_w^2/k_BT$ ($\s_w^2$ being the variance of the
Gaussian work distribution).  For quasi-reversible processes in the linear
response regime~\cite{GorRitBus03}, the fluctuation-dissipation theorem
implies $\s_w^2\simeq 2k_BTw^{\rm mp}_{\rm dis}$ giving $w^{\dag}_{\rm
dis}\simeq -w^{\rm mp}_{\rm dis}$, i.e. trajectories with negative
values of the dissipated work that are of the order (in absolute value) of the
average dissipated work must be sampled to quantitatively verify the
validity of the Jarzynski equality. An example of such a quasi-reversible
process, where $P_N(w_{\rm dis})$ is exactly a Gaussian, is the case of
a Brownian particle subjected to an harmonic potential and dragged in a
fluid~\cite{WanSevMitSeaEva02,MazJar99,ZohCoh03}.

\section{Numerical solution of the equations}
\label{numerics}
Equations \eqqq{a11a}{a11b}{a11c} can be numerically solved in general. 
We assume Glauber transition rates given by
\be
p^{\rm up}(H)=p^{\rm tot}(H)q(H)~~~;~~~p^{\rm down}(H)=p^{\rm
tot}(H)(1-q(H))
\label{num1}
\ee
with $q(H)=(1+\tanh(\beta\mu H))/2$ and $p^{\rm
tot}(H)=1/\tau_{\rm relax}(H)=\alpha(H)$ corresponding to the inverse of the
relaxation time. In this case,
\bea
p^{\rm up}(s)=\alpha(H(s)) \frac{\exp(\beta\mu H(s))}{2\cosh(\beta\mu H(s))}\label{num2a}\\
p^{\rm down}(s)=\alpha(H(s)) \frac{\exp(-\beta\mu H(s))}{2\cosh(\beta\mu
H(s))}~~~~.
\label{num2b}
\eea
Inserting these expressions in \eqq{a10a}{a10b} we obtain,
\bea
c(s)= -\alpha(H(s))\frac{\sinh(2\gamma(s)+\beta\mu H(s))}{\cosh(\beta\mu
H(s))}+\alpha(H(s))\tanh(\beta\mu H(s))
\label{num3a}\\
d(s)=\alpha(H(s))\frac{\cosh(2\gamma(s)+\beta\mu H(s))}{\cosh(\beta\mu
H(s))}-\alpha(H(s))~~~.
\label{num3b}
\eea

The solution of the equations consists of the
following steps:

\begin{enumerate}
\item{\bf Solution of ${\bf \gamma_{\lambda}(s)}$.} 
With the boundary condition at the final time $s=t$,
 $\gamma_{\lambda}(t)=0$, Eq.~\eq{a11c} has to be numerically integrated
 backwards in time. Inserting \eqq{num2a}{num2b} in \eq{a11c} we obtain,
\be
\dot{\gamma}(s)=\lambda\mu\dot{H}(s)+\alpha(H(s))\sinh(\gamma(s))\bigl(\cosh(\gamma(s))
+\sinh(\gamma(s))\tanh(\beta\mu H(s)) \bigr)~~~~.
\label{num4}
\ee
However, a direct numerical integration of
this equation leads to divergences and numerical instabilities. It is
then convenient to express \eq{num4} in terms of a new variable
$\eps(s)=1/\cosh(\gamma(s))$ which displays smooth behavior. Equation
(\ref{num4}) becomes,
\bea
\dot{\eps}(s)=-\frac{\tanh(\gamma(s))}{\cosh(s)}\Bigl(\lambda\mu\dot{H}(s)+\nonumber\\
\alpha(H(s))\sinh(\gamma(s))\bigl(
\frac{1}{\eps(s)}+\sinh(\gamma(s))\tanh(\beta\mu H(s)) \bigr)  \Bigr)
\label{num5}
\eea
with the boundary condition $\eps(t)=1$. This equation can then be easily
numerically integrated to give $\gamma_{\lambda}(s)$  for a given value of $\lambda$.

\item{\bf Solution of ${\bf m_{\lambda}(s)}$.} 
Once the solution of \eq{num5} for a given value of $\lambda$,
$\gamma_{\lambda}(s)$,  is found, then
it is possible to integrate \eq{a11b} to find $m_{\lambda}(s)$. Because
\eq{a11b} is linear its
solution can be explicitly written, 
\be
m_{\lambda}(s)=m_{\lambda}(0)\exp\bigl(\int_0^sA_1(u)du\bigr)+\int_0^sdu
A_2(u)\exp\bigl(\int_u^sA_1(v)dv\bigr)
\label{num6}
\ee
with the definitions,
\bea
A_1(s)=\alpha(H(s))\frac{\sinh(2\gamma_{\lambda}(s)+\beta\mu
H(s))}{\cosh(\beta\mu H(s))}\label{num7a}\\
A_2(s)=-\alpha(H(s))\frac{\cosh(2\gamma_{\lambda}(s)+\beta\mu
H(s))}{\cosh(\beta\mu H(s))}\label{num7b}
\eea
with the initial condition
$m_{\lambda}(0)=\tanh(\gamma_{\lambda}(0)+\beta\mu H_i)$. 

\item{{\bf Evaluation of ${\bf w,s(w),{\cal F}(w)}$}.}
Once $\gamma_{\lambda}(s),m_{\lambda}(s)$ are known then we can evaluate 
$w$ using \eq{a11a}, the entropy
$s(w)$ using \eq{as} and the free energy ${\cal F}(w)=w-k_B T
s(w)$. 

\item{\bf Dependence of the numerical algorithm on the sign of ${\bf \lambda}$.}
We must emphasize that the solution of the equations previously
described only works in a sector of values of $\lambda$ of a given sign,
$\lambda<0$, and leads to numerical instabilities in the other sector,
$\lambda>0$, indicative that the transformation
$\eps(s)=1/\cosh(\gamma(s))$ is inappropriate for $\lambda>0$. We have
found a simple way out to this problem. It can be easily proven that the
solution of \eq{a11c} for a given value of $\lambda>0$ is equivalent to
the solution of that equation with the value of $\lambda$ with its sign
reversed ($-\lambda<0$) and for the reversed field protocol
$H_r(s)=-H(s)$ (the subindex $r$ stands for reversed).  Eq.~(\ref{a11c})
can then be solved and the resulting reversed solutions
$m_r(s),\gamma_r(s),c_r(s),d_r(s)$ give the final solutions for the
original value of $\lambda>0$:
$m(s)=-m_r(s),\gamma(s)=-\gamma_r(s),c(s)=-c_r(s),d(s)=d_r(s)$ (all
change sign except $d(s)$). At first glance, this symmetry property
might seem to be related to the content of the fluctuation theorem. However
this relation is only apparent because the reversed process in this case
does not correspond to the time-reversal protocol which should be
instead $H_r(s)=H(t-s)$ (see the discussion below in Sec.~\ref{ft}).

\end{enumerate}

\begin{figure}
\begin{center}
\epsfig{file=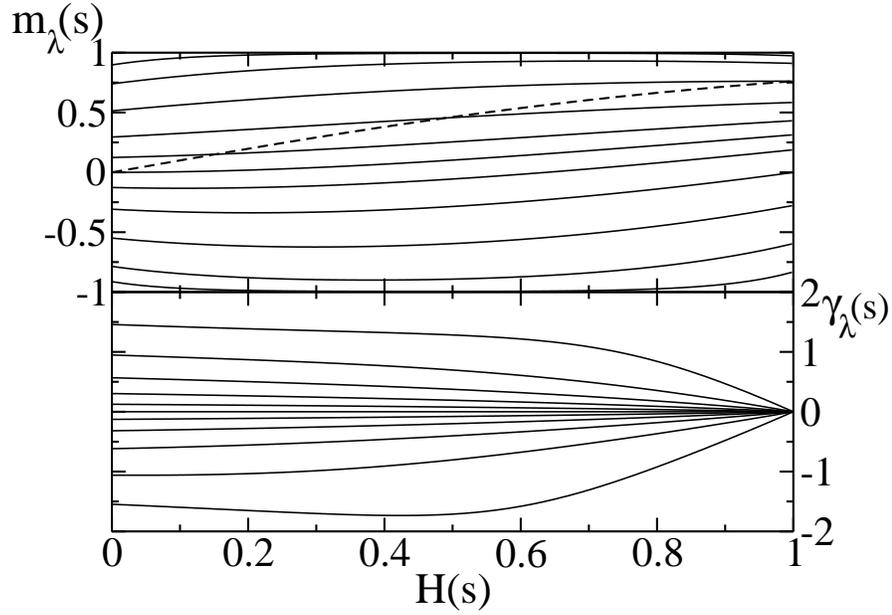,angle=-90, width=11.8cm}
\end{center}
\caption{\small Protocol with $H_i=0,H_f=1$ and ramping speed $r=1$. Curves
correspond to different values of $\lambda$
($\lambda=-5,-2,-1,-0.5,-0.2,0.,0.2,0.5,1,2,5$ from top to bottom in the
upper and lower panel). Upper panel: Magnetization $m_{\lambda}(s)$
obtained from \eq{num6}. The dashed line is the equilibrium
magnetization $m_{\rm eq}(H)=\tanh(H)$ corresponding to the reversible
ramping experiment $r=0$. Lower panel: Lagrange multiplier
$\gamma_{\lambda}(s)$ obtained from \eq{num5}. Note the boundary condition
$\gamma_{\lambda}(t)=0$ and the presence of the most probable trajectory
$\gamma_{\lambda=0}(s)=0,\forall s$.}
\label{fig2}
\end{figure}

\begin{figure}
\begin{center}
\epsfig{file=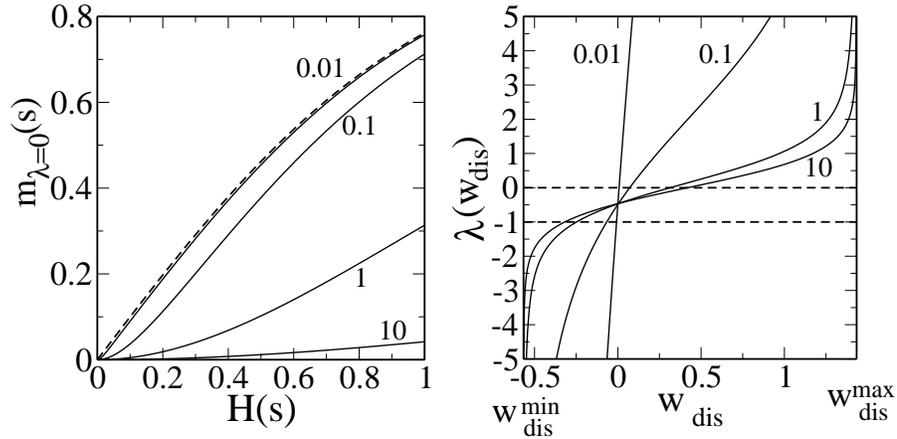,angle=-90, width=11.8cm}
\end{center}
\caption{\small Protocol with $H_i=0,H_f=1$ and different ramping speeds
$r=0.01,0.1,1,10$ (indicated by numbers along the continuous curves in both panels). The reversible
work is $w_{\rm rev}=-0.433781$ and $w_{\rm max}=1,w_{\rm dis}^{\rm
max}=w_{\rm max}-w_{\rm rev}=1.433781,w_{\rm dis}^{\rm
min}=w_{\rm min}-w_{\rm rev}=-0.566219$. Left panel:
magnetization evolution for the most probable trajectories. The dashed
line corresponds to the reversible trajectory for $r\to 0$. Right panel:
Lagrange multiplier $\lambda(w_{\rm dis})$ for different ramping speeds. The intersection
of the different curves with the dashed line $\lambda=0$ gives $w^{\rm
mp}$ while the intersection with $\lambda=-1$ gives $w^{\dag}$. The
intersection of all lines at different speeds around $\lambda=-0.5$ is
only accidental (looking at a larger resolution or considering other
parameters for the protocol such common crossing does not exist).}
\label{fig3}
\end{figure}

\begin{figure}
\begin{center}
\epsfig{file=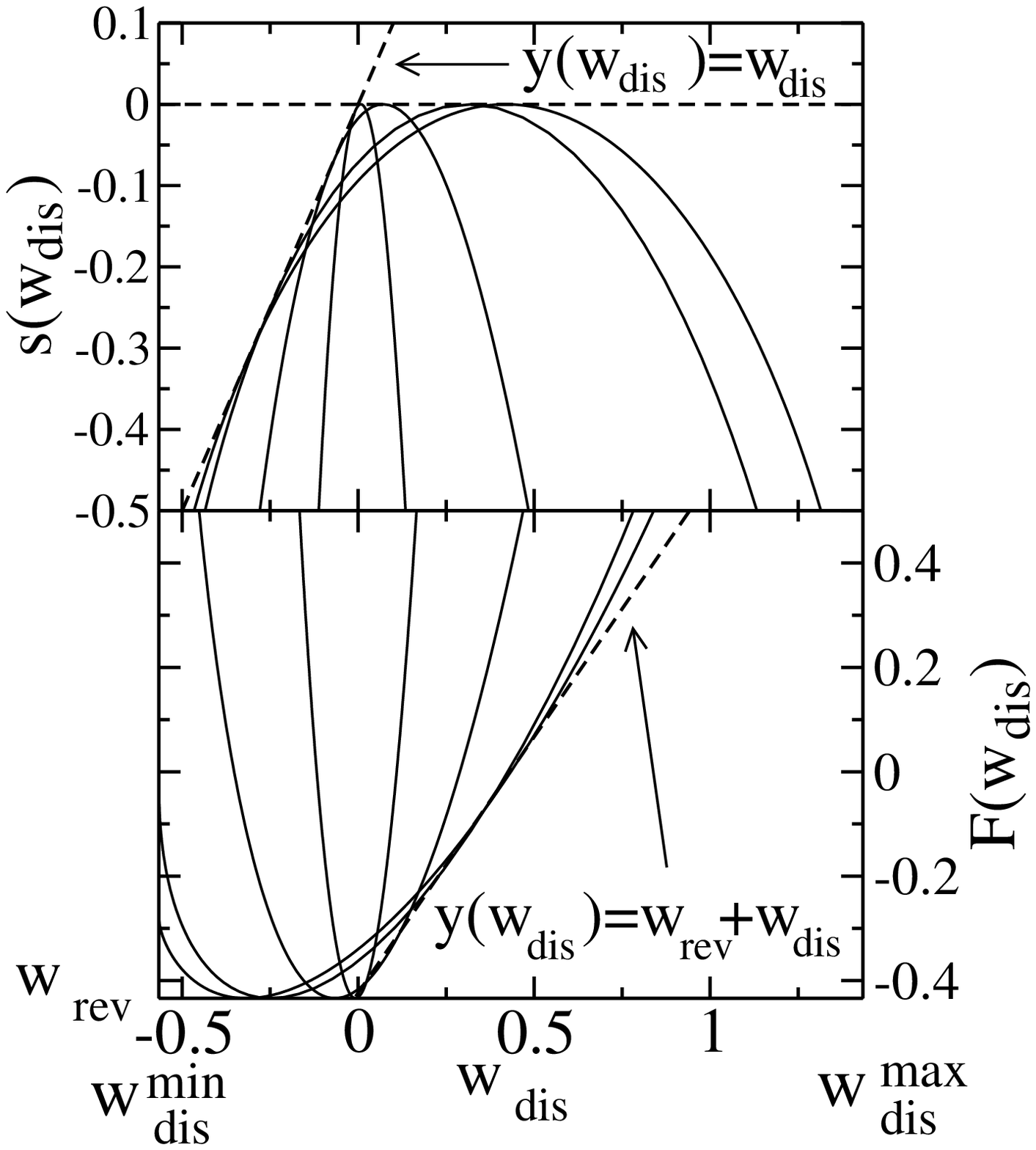,angle=0, width=14cm}
\end{center}
\caption{\small Same parameters and ramping speeds as in
Figure~\ref{fig3}. Narrower curves correspond to lower ramping
speeds. Upper panel: Dynamical entropies $s(w)$ plotted as functions of
$w_{\rm dis}=w-w_{\rm rev}$. According to the trajectory thermodynamics
relations \eqq{eqmicro1}{eqmicro2} the straight line $y(w_{\rm
dis})=w_{\rm dis}/k_BT$ (we take $k_BT=1$) is tangent to the curve
$s(w_{\rm dis})$ at $w_{\rm dis}=w_{\rm dis}^{\dag}=w^{\dag}-w_{\rm
rev}$ and crosses the $w_{\rm dis}$-axis at $w_{\rm dis}=0,s(w_{\rm
dis})=0$. All entropies vanish at $w_{\rm dis}=w_{\rm dis}^{\rm
mp}=w^{\rm mp}-w_{\rm rev}$. Lower panel: Trajectory free-energy ${\cal
F}(w_{\rm dis})$. It is identical to the equilibrium free-energy change
$\Delta F=w_{\rm rev}$ at $w_{\rm dis}=w_{\rm dis}^{\dag}$. According to
the same relations \eqq{eqmicro1}{eqmicro2} the straight line $y(w_{\rm
dis})=w_{\rm rev}+w_{\rm dis}$ is tangent to the curve ${\cal F}(w_{\rm
dis})$ at $w_{\rm dis}=w_{\rm dis}^{\rm mp}=w^{\rm mp}-w_{\rm rev}$ and
crosses the $w_{\rm dis}$-axis at $w_{\rm dis}=0,{\cal F}(w_{\rm
dis})=w_{\rm rev}$.}

\label{fig4}
\end{figure}

For the present numerical analysis, and for the sake of simplicity, we will
consider a particular example where the ramping field $H(s)$ changes
from $H(0)=H_i$ to $H(t)=H_f$ at a constant rate $r=\dot{H}$,
\be
H(s)=H_i+rs~~~,~~~r=\dot{H}=\frac{H_f-H_i}{t}~~~.
\label{num0}
\ee
We will also consider $p^{\rm tot}(H)=\alpha$ independent of the
field. This tantamount to assume that $p^{\rm tot}(H)$ corresponds to a
microscopic attempt frequency or, rather, that the activation barrier is
field independent.  We numerically solved the equations in natural units
$\mu=k_B T=1$ and we have chosen $\alpha=1$ as the characteristic
relaxation timescale of the system.  Results for different values of
$H_i,H_f$ have been obtained by doing ramping experiments at different
values of the ramping speed $r$.  In Figure~\ref{fig2} we show, in the
particular example $H_i=0,H_f=1$, the results for the magnetization
trajectory solutions $m_{\lambda}(s)$ and the Lagrange multiplier
$\gamma_{\lambda}(s)$. These are plotted as a function of the
time-dependent field $H(s)$ for different values of $\lambda$ and for a
given value of the ramping speed. In Figs.~\ref{fig3},~\ref{fig4} we
show several trajectory thermodynamics quantities at different ramping
speeds ($r=0.01,0.1,1,10$). In the left panel of Figure~\ref{fig3} we
plot the magnetization for the most probable trajectories
$m_{\lambda=0}(s)$ as a function of $H(s)$. In the right panel of
Figure~\ref{fig3} and in Figure~\ref{fig4} we show the different
trajectory thermodynamics quantities as a function of $w_{\rm dis}$: the
inverse temperature $\lambda(w_{\rm dis})$, the trajectory entropy
$s(w_{\rm dis})$ and the trajectory free-energy ${\cal F}(w_{\rm dis})$.


\subsection{Average and variance of the work distribution}
\label{mp}
\begin{figure}
\begin{center}
\epsfig{file=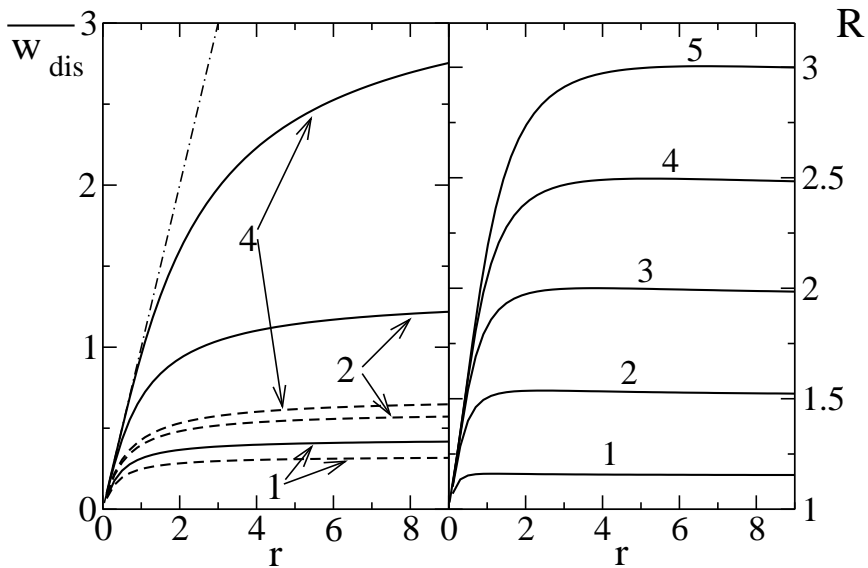,angle=-90, width=12cm}
\end{center}
\caption{\small Ramping experiments with $H_i=0$ and different values of $H_f$
as function of the ramping speed $r$. Left panel: $w_{\rm dis}$
(continuous lines) and $-w^{\dag}_{\rm dis}$ (dashed lines) for
different values of $H_f$ indicated in the figure. The dash-dotted
straight line corresponds to the linear response behavior $w_{\rm
dis}=r$ given by \eq{mp6}. Note that $w^{\dag}_{\rm dis}$ is negative so
we changed its sign in order to compare it with $w_{\rm dis}$.  Right
panel: Fluctuation-dissipation ratio $R$ as a function of $r$ evaluated
at different values of $H_f$ (from bottom to top, $H_f=1,2,3,4,5$).}
\label{fig5ab}
\end{figure}

As has been schematically depicted in Figure~\ref{fig1} there are
different work quantities that can be of relevance to characterize work
fluctuations. We have already defined the most probable work $w^{\rm
mp}$ and the work $w^{\dag}$. Another important quantity is the average
work $\overline{w}$,
\be
\overline{w}=\int_{w_{\rm min}}^{w_{\rm max}}w\,P_N(w)\,dw
\label{mp0}
\ee
where $P_N(w)$ was defined in \eq{a3} or in approximate form in
\eq{distrib}. In most cases (for instance, when the work distribution
has asymmetric tails) the average work $\overline{w}$ is different from
the most probable work $w^{\rm mp}$.  $\overline{w}$ can be lower or
higher than the most probable work $w^{\rm mp}$. However, in our
large-$N$ theory, $w^{\rm mp}=\overline{w}$ and we will use indistinctly
both quantities in this section. We defer the discussion about
finite-size effects in these quantities until Sec.~\ref{fse}.  Another
important quantity that characterizes the work distribution is its
variance,
\be
\sigma_w^2=\overline{w^2}-(\overline{w})^2=\overline{w_{\rm
dis}^2}-(\overline{w_{\rm dis}})^2~~~.
\label{mp1}
\ee
The average work $\overline{w}$ (or $w^{\rm mp}$) is the most relevant
physical quantity that connects with classical thermodynamics.  The
second law of thermodynamics establishes that it cannot be lower than
the reversible work, $\overline{w}\ge w_{\rm rev}$. However, it is clear
that there can be WF such that $w<w_{\rm rev}$. These have
been called transient violations (TV) of the second law. The relevant
work value characterizing this sector of trajectories is given by
$w^{\dag}$. Clearly, $\overline{w}$ is always higher
than $w^{\dag}$. In Figure~\ref{fig5ab} (left panel) we show the dependence
of $w_{\rm dis},w^{\dag}$ with the ramping speed when the field is
ramped from $H_i=0$ to $H_f$ for different values of $H_f$.
It is possible to write down explicit analytic expressions for the cumulants of the
distribution $P_N(w)$ in the large-$N$ limit. Interestingly, and due to
the non-interacting character of the model \eq{a1}, the cumulants derived from
the large-$N$ approach are exact at all values of $N$, see Sec.~\ref{fse}.
In particular, the first moment is given by,
\bea
\overline{w_{\rm dis}}=w^{\rm mp}-w_{\rm
rev}=\nonumber\\
2\mu\int_0^tds\dot{H}(s)\int_0^sdu\dot{H}(u)\frac{\partial
q(H)}{\partial H}\Bigr|_{H=H(u)}\exp\Bigl(-\int_u^s dv \alpha(H(v))\Bigr)~~~.
\label{mp2}
\eea
The expression for the second cumulant or variance can be obtained by expanding the function $s(w)$
\eq{as} up to second order with $\lambda(w)$ as the small parameter. Using the
result $s'(w^{\rm mp})=0$ we get 
\be 
s(w)=s(w^{\rm mp})+\frac{1}{2}\Bigl (\frac{\partial^2 s(w)}{\partial w^2}\Bigr)(w-w^{\rm mp})^2+{\cal
O}\Bigl[(w-w^{\rm mp})^3\Bigr]~~~.
\label{mp3}
\ee
From \eq{distrib} and \eq{maximum} we obtain the relation,
\be
\sigma_w^2=\frac{1}{N}\Bigl[\frac{\partial^2 s(w)}{\partial
w^2}\Bigr|_{w=w^{\rm mp}} \Bigr]^{-1}=\frac{1}{N}\frac{dw(\lambda)}{d\lambda}\Bigr|_{\lambda=0}
\label{mp4}
\ee
We do not reproduce the details of this lengthy calculation here, the
same results have been already obtained in a slightly different context
in a previous work and in the limit of large free-energy changes $\Delta
F$ as compared to $k_BT$~\cite{RitBusTin02}.

\begin{figure}
\begin{center}
\epsfig{file=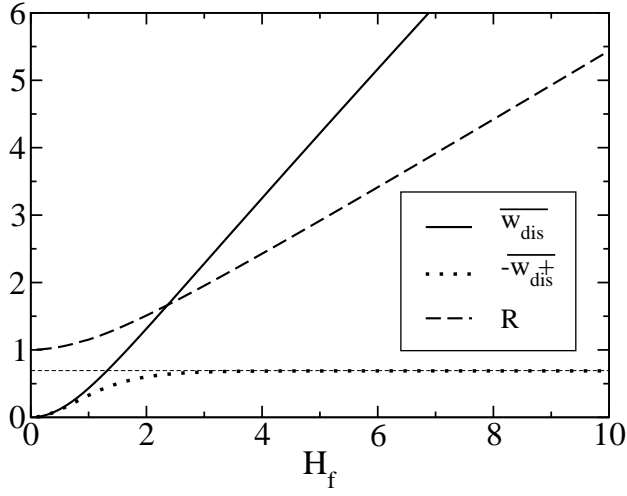,angle=-90, width=10cm}
\end{center}
\caption{\small Ramping experiment with $H_i=0$ and $r=100$ as a function of
$H_f$. The average dissipated work $w_{\rm dis}^{\rm mp}$ (continuous
line) and $R$ (dashed line)
increase with the field but $w_{\rm dis}^{\dag}$ (dotted line) saturates
to a finite value equal to $-\log(2)$ (we represent $-w_{\rm
dis}^{\dag}$ in order to compare with  $w_{\rm dis}^{\rm mp}$).}
\label{fig5c}
\end{figure}

Another interesting aspect of the present theory is that it is possible
to expand the cumulants around the limits $r\to 0$ or
$r\to\infty$. The former is particularly interesting because it
corresponds to the so-called linear-response regime. 
In Reference \cite{RitBusTin02} this regime was considered
by expanding the average dissipated work up to linear order in the
perturbation speed. By using dimensional
arguments and direct comparison with the equivalent expression derived in the
context of mechanical force~\cite{RitBusTin02} we can derive the
following result,
\be
\overline{w_{\rm dis}}=\frac{\mu\Delta M}{N\tau_{\rm relax}(H_c=0)}r+{\cal O}(r^2)
\label{mp6}
\ee
where $\Delta M=M_{\rm eq}(H_f)-M_{\rm eq}(H_i)$ is the difference of
equilibrium magnetizations between the initial and final values of the
field whereas $\tau_{\rm relax}(H_c=0)=1/p_{\rm tot}(H_c=0)$ is the
relaxation time at the critical value of the field where the
configurations $\s=+1$ and $\s=-1$ are equiprobable (i.e. $H_c=0$).  The
linear response regime breaks down for large ramping speeds when
$\overline{w_{\rm dis}}\ll w_{\rm rev}$. An interesting quantity
quantifying deviations from the linear-response regime is the
fluctuation-dissipation ratio $R$ defined by,
\be
R=\frac{\sigma_W^2}{2k_B T \overline{W_{\rm dis}}}=\frac{\sigma_w^2}{2k_B T \overline{w_{\rm dis}}}~~~.
\label{mp5}
\ee
In the limit of small $r\to 0$, when $\overline{w_{\rm dis}}\propto r$,
then $R$ converges to 1 (in agreement with the fluctuation-dissipation
theorem, a result that in the context of steady-state systems has been
proven in~\cite{GalCoh95}) but deviates from 1 as $r$ increases. In the
right panel of Figure~\ref{fig5ab} we show $R(r)$ when the field is
ramped from $H_i=0$ to $H_f$ at different values of $H_f$. In this case
the behavior of both $\overline{w_{\rm dis}}$ and $R$ is monotonic with
$r$.  In Figure~\ref{fig5c} we show the same ramping experiments but
comparing, for a given ramping speed, the results for $w_{\rm
dis},-w^{\dag},R$ as a function of $H_f$ . For values of $H_f$ small
enough the ramping process is well described by the linear
response-approximation discussed in Sec.~\ref{thermo} and $w^{\dag}_{\rm
dis}\simeq -w^{\rm mp}_{\rm dis}$.

\begin{figure}
\begin{center}
\epsfig{file=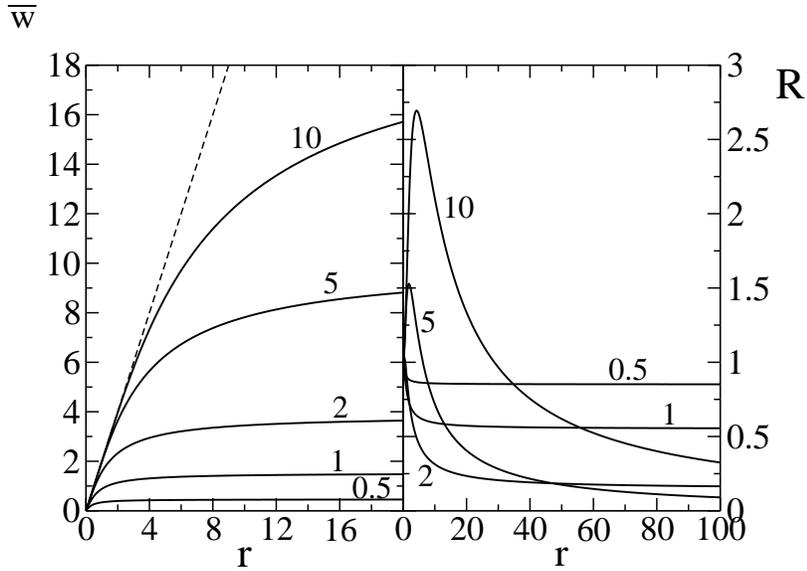,angle=-90, width=11cm}
\end{center}
\caption{\small Average work $w$ (left) and fluctuation-dissipation
ratio $R$ (right) for the case $H_f=-H_i=H_A$ as a function of the
ramping speed. The different curves correspond to different values of
the amplitude field $H_A$. These are indicated in the plot besides
each curve. The straight dashed line in the left panel corresponds to
the linear-response relation $\overline{w}=2r$ in \eq{mp6}. As
explained in last paragraph of Sec.~\ref{ft}, and for this particular
protocol, the relation $w^{\dag}=-w^{\rm mp}$ is exact for all values
of $r$ and $H_A$. Moreover, for large values of $H_A$ the work
coincides with the heat exchanged, see Sec.~\ref{heat}.}
\label{fig7ab}
\end{figure}

Let us finish this section emphasizing that the dependence $R(r)$ can
be quite complicated and even non-monotonic in some cases. Such
behavior is observed in the case where the ramping protocol is given
by $H(s)=H_A(1-2s/t)$, i.e. the field starts at a given value
$H_i=-H_A$ ($H_A$ denotes the field amplitude) and increases until its
reversed value $H_f=H_A$ is reached. This case is of much interest
regarding heat fluctuations and is discussed in detail in
Sec.~\ref{heat}. In Figure~\ref{fig7ab} we show the behavior of the
average work $\overline{w}$ (equal to the average dissipated work as
$w_{\rm rev}=0$ due to the independence of the free energy on the sign
of $H_A$) and $R$ as a function of the ramping speed for different
values of $H_A$.
\section{The fluctuation theorem}
\label{ft}
Saddle-point equations \eqqq{a11a}{a11b}{a11c} were derived in the
large-$N$ limit. Indeed \eq{distrib} is not exact for finite $N$ but has
corrections. However, the results obtained for $s(w)$ exactly satisfy
the fluctuation theorem of Crooks~\cite{Crooks98}. This theorem states
the following: let us consider a process where the system is perturbed
according to a protocol $H_F(s)$ during the time interval $[0,t]$,
initially the system being in equilibrium at the value of the field
$H_i=H_F(0)$. We will call this the forward (F) process. Let us now
consider the reverse process defined as the time-reversed of the forward
process: in this process the system starts in equilibrium at time $s=0$
at the value of the field $H_i=H_F(t)$ and the field is changed
according to the protocol $H_R(s)=H_F(t-s)$. Let the distribution of
works generated in this way be $P_F(W),P_R(W)$ for the forward (F) and
reverse (R) processes respectively. The two distributions satisfy the
following relation \cite{Crooks98},
\be
\frac{P_F(W)}{P_R(-W)}=\exp\Bigl(\frac{W-\Delta
F}{k_BT}\Bigr)=\exp\Bigl(\frac{W_{\rm dis}}{k_BT}   \Bigr)
\label{ft1}
\ee
where $\Delta F=F(H_f)-F(H_i)$ is the change in the equilibrium
free-energy. By rewriting this identity as
$P_R(-W)=P_F(W)\exp\Bigl(\frac{-W+\Delta F}{k_BT}\Bigr)$ and integrating
it between $W=-\infty$ and $W=\infty$ leads to the Jarzynski equality
$<\exp(-W_{\rm dis}/k_BT)>_F=1$ where $<...>_F$ stands for a dynamical
average over work values obtained along the forward process.

If we now substitute \eq{distrib} into the relation \eq{ft1} we obtain,
\be
s_F(w_{\rm dis})-s_R(-w_{\rm dis})=\frac{w_{\rm dis}}{k_BT}
\label{ft2}
\ee
where we have taken $P_{F,R}(W)\propto \exp(Ns_{F,R}(w))$ and we have
disregarded the normalization constant in the distribution \eq{distrib}
as unimportant. Because the quantity $s(w)$ used in \eq{distrib} is only
exact in the large-$N$ limit one might be tempted to think that \eq{ft2}
does not hold. To prove the validity of \eq{ft2} we rewrite \eq{ft1} in the
following way,
\be
\frac{1}{N}\log(P_F(W))-\frac{1}{N}\log(P_R(-W))=\frac{W_{\rm dis}}{Nk_BT}~~~.
\label{ft3a}
\ee
In the large-$N$ limit the distributions \eq{distrib} satisfy,
\be
\lim_{N\to\infty}\frac{1}{N}\log(P_{F,R}(W))=s_{F,R}(W)
\label{ft4a}
\ee
and therefore \eq{ft2} is exact with $w_{\rm rev}=\lim_{N\to
\infty}\Delta F/N$. The present approach seems quite general so the
trajectory entropy derived in a large-$N$ theory in any statistical
model (interacting or non-interacting) must verify the relation
\eq{ft2}.  Another interesting relation that can be obtained from \eq{ft2}
relates the values of $w^{\rm mp},w^{\dag}$ for the forward and
reverse processes. Differentiating \eq{ft2} respect to $w$ we obtain,
\be
s'_F(w)+s'_R(-w)=\lambda_R(-w)-\lambda_F(w)=\frac{1}{k_BT}
\label{ft4}
\ee
where we used \eq{maximum}. Therefore, the identity $s'_F(w^{\rm
mp})=\lambda_F(w^{\rm mp})=0$ \eq{maximum2} implies $s'_R(-w^{\rm
mp})=\lambda_R(-w^{\rm mp})=1/k_BT$. From \eq{eqmicro1} we then infer
that $w^{\dag}$ for the reverse process is identical to $-w^{\rm mp}$
for the forward process and vice versa. This relation is quite
interesting because it suggests that in order to estimate (e.g. in
experiments) the value of $w^{\dag}$ for a given non-equilibrium process
it is enough to determine $w^{\rm mp}$ for the reversed process, a
quantity that is experimentally accessible.

An interesting case of \eq{ft1} occurs whenever the forward and reverse
processes are symmetrical mirror images,
$H_F(s)=-H_R(s)=-H_F(t-s)$. This can be accomplished when $H_A=H_f=-H_i$
along the forward process and the protocol satisfies $H(s)=-H(t-s)$. In
this case the forward and the reverse work distributions are identical,
$W_{\rm rev}=\Delta F=0$ (or $w=w_{\rm dis}$) and \eq{ft2} reads,
\be
s(w)-s(-w)=\frac{w}{k_BT}
\label{ft3}
\ee
where we have used $s(w)=s_{F}(w)=s_{R}(w)$.  The validity of \eq{ft3}
can be further demonstrated by close inspection of equations
\eqqq{a11a}{a11b}{a11c}. Let $s(w)$ be the value of the dynamical
entropy for a given value of the work $w$ associated to the value of
the Lagrange multiplier $\lambda$ and the magnetization $m(s)$. Then,
for the reversed value of the work $-w$, it is possible to show that
the corresponding solutions are: $-\lambda-1/k_BT$ for the Lagrange
multiplier and $-m(t-s)$ for the magnetization solution. The resulting
entropy is then $s(-w)=s(w)-w/k_BT$ as given in \eq{ft3}. A remarkable
consequence of this special case is the aforementioned fact that
$w^{\dag}=-w^{\rm mp}$ at any ramping speed and for any value of ther
amplitude of the field $H_A$. This case was already shown in
Figure~\ref{fig7ab}.  The present symmetric case is specially
interesting because the work done upon the system can be identified
with the heat exchanged between the system and the bath. The
conditions required for such identification are discussed next.

\section{Heat fluctuations and tails}
\label{heat}
Until now we focused our efforts to investigate work
fluctuations. However, in the same way as the work fluctuates, the heat
exchanged between the system and the bath also does.  The validity of the
mechanical equivalence of heat (the content of the first law of
thermodynamics) suggests that there should not be an important
difference between heat and work. Heat is more difficult to
experimentally measure than work and this is the reason why we 
tend to be more interested in the former.

A motivation to investigate heat fluctuations has recently arisen in the
context of steady state and aging systems. In the first case, heat
fluctuations were investigated for the simple model of a bead dragged
through a viscous fluid \cite{ZohCoh03}. In the second case these were
studied for the case of a spin-glass model characterized by slow
dynamics and aging \cite{CriRit04,Ritort03,Ritort04}. In both cases, a
Gaussian component was identified in the heat distributions together
with some exponential tails. For the steady state system these
exponential tails were consequence of the validity of an asymptotic
fluctuation-theorem for the heat while in the aging system the tails
were the signature of intermittency effects that have been
experimentally observed in glasses and
colloids~\cite{BelCilLar01,Cip03}.

\begin{figure}
\begin{center}
\epsfig{file=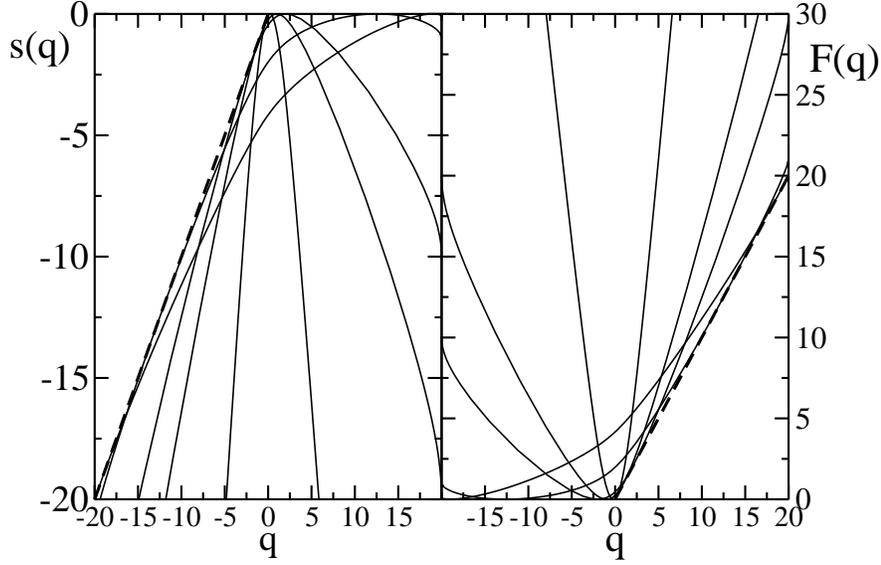,angle=-90, width=12cm}
\end{center}
\caption{\small Trajectory entropy $s(q)$ and trajectory free-energy ${\cal
F}(q)$ for the case $H_f=-H_i=10$ at ramping speeds $r=0.1,0.5,1,10,100$
(from the most narrower to the most wider distributions). The dashed
line in the left panel is $y(q)=q/k_BT$ (we take $k_BT=1$) and is
tangent to $s(q)$ at a value $q^{\dag}$. The dashed line in the right
panel corresponds to $y(q)=q$ and is tangent to ${\cal F}(q)$ at the
value $q^{\rm mp}$. Both $q^{\rm mp},q^{\dag}$ depend on the ramping
speed.}
\label{fig6ab}
\end{figure}

\begin{figure}
\begin{center}
\epsfig{file=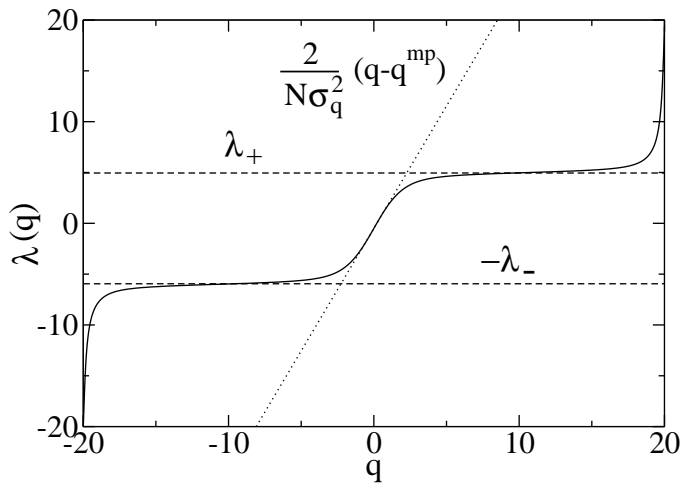,angle=-90, width=10cm}
\end{center}
\caption{\small $\lambda(q)$ for the same parameters as in Figure~\ref{fig6ab} in
the case $r=0.1$. We note the presence of a linear behavior for
$\lambda(q)$ for small values of $q$, $\lambda(q)=(2/N\s_{q}^2)(q-q^{\rm
mp})$ ($q^{\rm
mp}=0.207,N\sigma_q^2=0.83$) and two plateaus for $q>>1$ and $q<<-1$
($\lambda_+=4.95,\lambda_-=5.95$). The former gives rise to the Gaussian
component in the heat distribution describing the statistics of most
probable values. The latter gives rise to two exponential tails for the
distribution describing the statistics of rare events.}
\label{fig6ab2}
\end{figure}

The heat along a given trajectory can be inferred using energy
conservation, $-Q+W=\Delta E$. To extract the heat we just need to
know the change in energy $\Delta E$ between the final and initial
configurations as well as the work $W$. Here we adopt the sign
convention (contrary to that adopted in the introduction
Sec.~\ref{intro}) that positive heat corresponds to net heat delivered
by the system on the surroundings. A particular case where work and
heat fluctuations are identical is the case described in the preceding
section where $H_f=-H_i=H_A$. Due to time reversal symmetry $W_{\rm
rev}=\Delta F=0$. Now, if the field amplitude $H_A$ is large enough,
the difference in energy $\Delta E=-\mu\Delta (MH)$ is practically
zero so $Q=W$. For example, if $H_A=5$, then $\tanh(5)=0.99991$ (as
always we take $\beta=\mu=1$) so the initial equilibrium magnetization
is $M_{\rm eq}(H_A)=-N$. The final magnetization after ramping the
field to $H_A$ is again of order $N$ and therefore the fluctuations
from trajectory to trajectory in $\Delta E$ are negligible as compared
to the total work.  In Figure~\ref{fig6ab} we show the trajectory
entropy and free-energy for the case $H_A=10$. We have chosen to
represent variables in terms of heat per particle $q=Q/N$ rather than
work to give a view of what general shape we can expect from heat
distributions.  In terms of the heat we expect that the same
mathematical definitions and relations that we defined in the case of
work are also valid.  For instance, the heat entropy and the heat
free-energy are defined in the same way as we did for their work
counterparts just replacing $w$ by $q$, in particular ${\cal
F}(q)=q-k_BT s(q)$.  Also the equivalent of \eq{maximum} holds,
\be
\frac{\partial s(q)}{\partial q}=-\lambda(q)
\label{heat-1}
\ee
The most probable heat $q^{\rm mp}$ ($\lambda(q^{\rm mp})=0$) and the
quantity $q^{\dag}$ ($\lambda(q^{\rm mp})=-1$) can also
be defined. Moreover, a relation equivalent to \eq{ft3} is also expected
to hold for large enough values of $H_A$,
\be
s(q)-s(-q)=\frac{q}{k_BT}~~~.
\label{heat0}
\ee
An interesting aspect of the heat entropy $s(q)$ shown in the left panel
of Figure~\ref{fig6ab} is the presence of quadratic behavior for small
values of $q$ ($q\simeq 0$) together with a linear behavior in the tails
($|q|>>1$).  These characteristic features of the heat entropy $s(q)$
can be inferred by looking at $\lambda(q)$, shown in
Figure~\ref{fig6ab2}. That figure shows that $\lambda(q)$ is linear with
$q$ for $q\simeq 0$, giving a quadratic behavior for $s(q)$ at small
values of $q$. This linear shape in $\lambda(q)$ corresponds to a
Gaussian distribution for $P(q)=\exp(s(q))$. It also shows that for a
wide range of $|q|$ values there are two plateaus at $\lambda(q)\sim
\lambda_+,-\lambda_-$ for positive and negative values of $q$
respectively. These plateaus correspond to the exponential tails in the
distribution.
This behavior is quite generic at all ramping speeds, the distinction in
$\lambda(q)$ between both plateaus and the linear behavior at small $q$
becomes more clear as the speed decreases. In such conditions, $q^{\rm
mp}$ is not very large and the linear response approximation holds. The
Gaussian sector describes the statistics of small and most probable
fluctuations, the exponential tails describe rare events and large
deviations. In what follows we analyze the Gaussian and exponential
tails in more detail.

In the region where both $q,q^{\rm mp}$ are not too large we have, 
\be
s(q)=-\frac{1}{N\sigma_q^2}(q-q^{\rm mp})^2~~~~~~q,q^{\rm mp}<<1
\label{heat1}
\ee
Substituting this relation in \eq{heat0} we get,
\be
s(q)-s(-q)=\frac{2q^{\rm mp}}{N\sigma_q^2}q=\frac{q}{k_BT}
\label{heat1a}
\ee
implying
\be
\frac{N\sigma_q^2}{2q^{\rm mp}k_BT}=1
\label{heat1b}
\ee
This result shows that the fluctuation-dissipation ratio \eq{mp5} is
equal to 1 if heat fluctuations are restricted to the sector of $q$
small. Small fluctuations are a key assumption of linear-response theory
which also leads to \eq{heat1b}.
 
\begin{figure}
\begin{center}
\epsfig{file=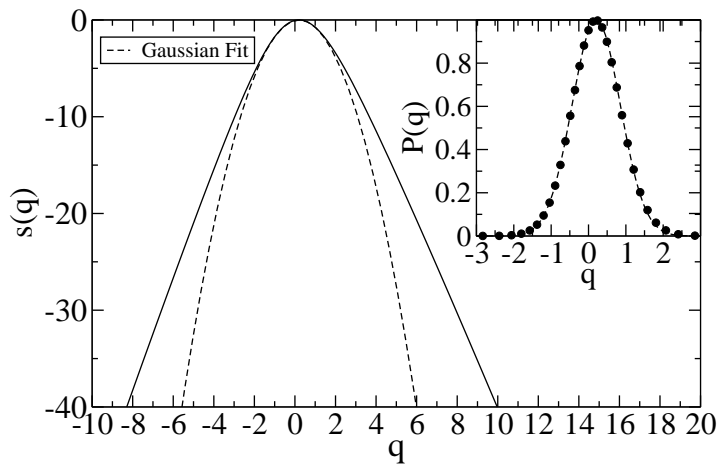,angle=-90, width=10cm}
\end{center}
\caption{\small Heat entropy $s(q)$ for the case $H_f=-H_i=10$ at ramping speed
$r=0.1$. Main figure: The sector of small or most probable fluctuations
$q\sim 0$ can be well fitted to the Gaussian \eq{heat1} (dashed line)
with parameters $q^{\rm mp}=0.207,N\sigma_q^2=0.83$ satisfying
\eq{heat1b}. The tails extend beyond the Gaussian central part and are
of exponential type as described in \eqq{heat2}{heat3} with
$\lambda_-=5.95,\lambda_+=4.95,C=9.16$. These exponential tails describe
the statistics of large deviations or rare events. Inset:
Heat-distribution $P(q)=\exp(s(q))$ (dots) and the Gaussian
fit (the dashed line of the main figure) showing that the small $q$
sector of fluctuations (those that are frequently observable)
is very well fitted by a Gaussian despite of the fact that rare-event
tails are big and observable only when plotting $s(q)$ or the
distribution $P(q)$ in logarithmic scale.}
\label{fig6cd}
\end{figure}

This quadratic behavior then goes over straight lines in the most
negative and positive sectors of $q$,
\bea
s(q)=C - \lambda_+ q~~~~~q>>1\label{heat2}\\
s(q)=C + \lambda_- q~~~~~q<<-1\label{heat3}
\eea
where $C$ is a constant and $\lambda_+,\lambda_-$ correspond to the
values of $\lambda(q)$ in the plateaus shown in
Figure~\ref{fig6ab2}. Note that the constant $C$ in \eqq{heat2}{heat3}
is the same in both sectors. In fact, the relation \eq{heat0} imposes
such constraint between the positive and negative tails in the
probability distributions. Substituting \eqq{heat2}{heat3} into
\eq{heat0} we obtain
\be
\lambda_--\lambda_+=\frac{1}{k_BT}
\label{heat4}
\ee
meaning that the width of the tails is larger for the negative values of
 $q$ than for positive values. This can be interpreted by saying that,
 despite of the fact that the average heat $q$ is positive, rare
 fluctuation events occur as often for $q<0$ (i.e. when the system adsorbs
 heat from the surroundings) as they do for $q>0$ (when the system delivers
 heat to the surroundings). The shape of the heat
 distribution $P(q)= \exp(s(q))$ is then dominated by a central
 Gaussian distribution with exponential tails at its extremes. These
 features are illustrated in Figure~\ref{fig6cd}.
If the amplitude field $H_A$ is not large enough there may be
contributions to the heat distribution coming out from the
fluctuations in the difference in energy between the initial and final
configurations. The effect of the value of $H_A$ on the value of the
average work and the fluctuation-dissipation ratio have been already
shown in Figure~\ref{fig7ab}, in particular non-monotonic behavior is
observed for $R$.

\section{Finite-size effects}
\label{fse}
The method we developed in this paper allowed us to calculate $P_N(w)$ in the
large-$N$ limit. However, due to the
non-interacting character of the model, all cumulants of the
distribution obtained in the large-$N$ limit are also exact for finite 
$N$. The proof is quite straightforward. Let us define the generation
function of all cumulants of the distribution $P_N(W)$ in \eq{a3},
\be
g_N(x)=\log\Bigl(\overline{\exp(xW)}\Bigr)=\log\Bigl( \int dW\exp(xW)P_N(W)\Bigr)~~~.
\label{fse1}
\ee
Cumulants of  $P_N(W)$ are obtained using the following formula,
\be
c_N(k)=\frac{\partial^k g_N(x)}{\partial x^k}\Bigr|_{x=0}~~~,
\label{fse2}
\ee
$k$ being a positive integer. Using the non-interacting character of the model then we can write,
\be
W=\sum_{i=1}^Nw_i  \rightarrow  g_N(x)=Ng_1(x)\rightarrow c_N(k)=Nc_1(k)
\label{fse3}
\ee
and therefore all cumulants of the distribution are independent of the
size of the system (except by a multiplicative constant equal to $N$).
This implies that the expression given for $\overline{w_{\rm dis}}$ in
\eq{mp2} and $R$ in \eq{mp5} are independent of $N$. Therefore, the
results we obtained in the large-$N$ limit are {\em exact for any finite
value of $N$}.

\begin{figure}
\begin{center}
\epsfig{file=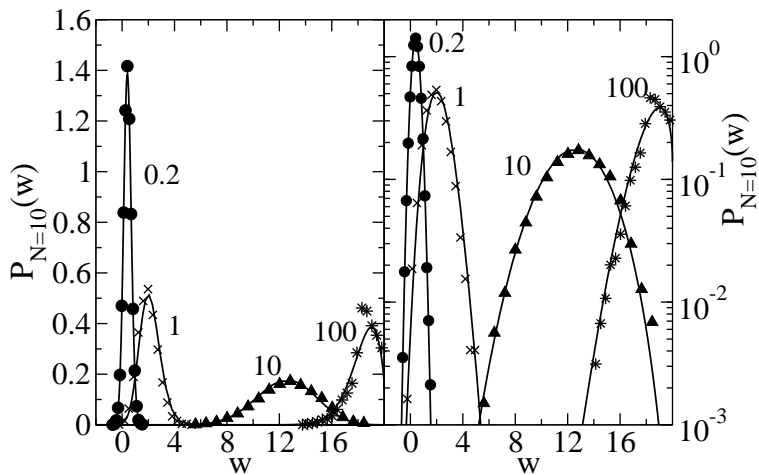,angle=-90, width=10cm}
\end{center}
\caption{\small Distributions $P_N(w)$ for case $H_f=-H_i=10$ and
$N=10$ at different ramping speeds (indicated along each curve). The
continuous lines are the results obtained from the present theory using
\eq{distrib}. The symbols are results obtained from numerical
simulations of the model for $10^4$ trajectories. The right panel is the
same figure but in logarithmic scale. For the largest ramping speed
$r=100$ there is a finite fraction of trajectories (about $37\%$ of the
total number of trajectories) where the spins have no time to
relax. These trajectories contribute with a singular term at $w=w_{\rm
max}=20$ to the distribution $P_{N=10}(w)$ as described in \eq{fs4}. It
cannot be captured by the present large-$N$ theory so we did not
include it in the histogram obtained from the numerical simulations.}
\label{fig9}
\end{figure}

However, albeit cumulants do not depend on $N$, the shape of the
distribution $P_N(w)$ in \eq{distrib} depends on the size $N$ and only
in the large-$N$ limit the approximate distribution \eq{distrib} becomes
exact. For instance, the value of $w^{\rm mp}$ depends on $N$ and
converges to $\overline{w_{\rm dis}}$ for large enough values of $N$. In practice,
already for $N=5-10$ convergence of the approximate distribution
\eq{distrib} to the exact result is excellent. In order to compare the
approximate distributions we obtain from our theory with the exact ones
at finite $N$ we have done numerical simulations of the model. The
simulation procedure is described below in Sec.~\ref{nano}. In
Figure~\ref{fig9} we show the distributions we have obtained for $N=10$
compared to the numerical simulations at different ramping speeds. The
agreement between theory (continuous lines) and simulations (symbols)
is good although it worsens progressively as the ramping speed increases
and the system is strongly driven out of equilibrium.  An important
feature of the distributions is observed for large $r$ and small $N$:
the presence of a finite fraction of trajectories that dissipate a
maximum amount of work equal to $w_{\rm max}=\mu(H_f-H_i)$. For these
trajectories the ramping speed is so high and the size so small that no
change in the initial configuration occurs along the trajectory. We will
call these {\em trapped} trajectories. The fraction of trapped
trajectories contributes with a term $\delta(w-w_{\rm max})$ to the work
distribution,
\be
P_N(w)=\tilde{P}_N(w)+\alpha(N)\delta(w-w_{\rm max})
\label{fs4}
\ee
where $\tilde{P}_N(w)$ is a continuous function and $\alpha(N)$ is a
size-dependent constant that decreases with $N$ and asymptotically
vanishes in the large-$N$ limit.  The delta function in \eq{fs4} is a
{\em small} $N$ contribution that is not captured by the present
large-$N$ theory. Nevertheless, it might be analytically derived using
the approach described below in Sec.~\ref{reconstruction}.

\begin{figure}
\begin{center}
\epsfig{file=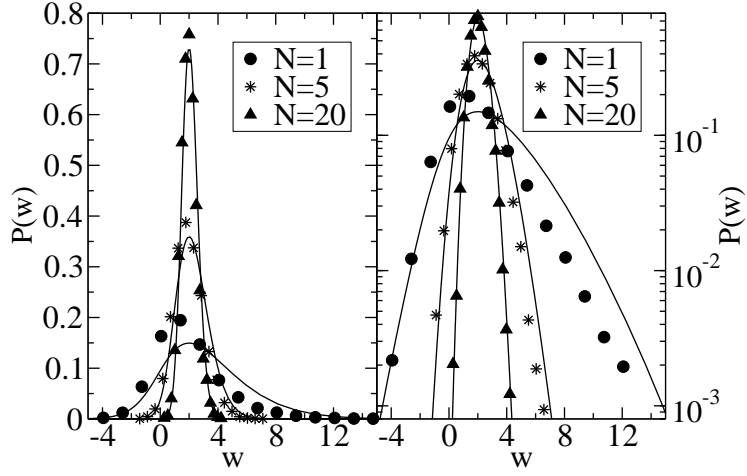,angle=-90, width=10cm}
\end{center}
\caption{\small The same as in Figure~\ref{fig9} but showing the dependence
of $P_N(w)$ with $N$ at $r=1$. The agreement is not good for $N=1$ in
the central region of the distribution but is reasonably good for the
left tail of the distribution (see the right plot in logarithmic
scale). The agreement improves noticeably beyond $N\simeq 5$.}
\label{fig10}
\end{figure}

\begin{figure}
\begin{center}
\epsfig{file=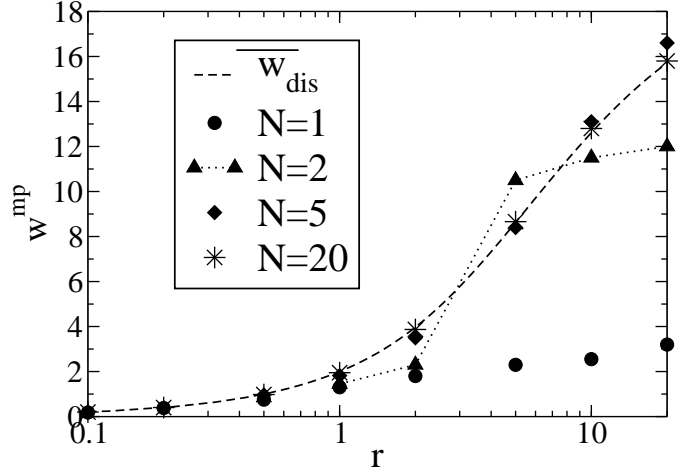,angle=-90, width=10cm}
\end{center}
\caption{\small The same parameters as in Figure~\ref{fig9} but now showing the
dependence of the most probable work value $w^{\rm mp}$ with $N$. The
different symbols in the points correspond to different sizes as
indicated in the legend of the figure. The continuous dashed line is
$w^{\rm mp}$ as derived from the theory in the large-$N$ limit (where
$w^{\rm mp}= \overline{w_{\rm dis}}$, the latter being independent of
$N$). In the linear-response
regime, $r<<1$, data have converged to the theory for all sizes. Although
finite-size effects are large for $N=1,2$ and $r>>1$, already for $N=5$
the simulations have converged to the theory at all ramping speeds. Data
for $N=2$ have been connected by a dotted line to emphasize the sharp
increase of $w^{\rm mp}$ around $r\simeq 5$. This sharp increase originates
from the presence of two separated peaks in the work distribution which
height coincide at a given value of the ramping speed around $r\simeq
5$.}
\label{fig11}
\end{figure}

In Figure~\ref{fig10} we show the effect of the size on the
distributions at a moderate ramping speed.  For $N=1$ the agreement is
not good although the behavior of the left tails is reasonably well
reproduced. However, already for $N=5$ the agreement has improved
considerably. We conclude that it is between $N=1$ and $N=5$ that
finite-size effects are important. In Figure~\ref{fig11} we confirm this
strong {\em small-$N$} dependence by plotting the most probable work as obtained from
the numerical simulations as a function
of $r$ for different sizes $N=1,2,5,20$. 

\subsection{Reconstructing $P_1(w)$ from the large-$N$ theory}
\label{reconstruction}
A crucial aspect of the present model is that it is
non-interacting. Therefore, if we were to know the work probability
distribution for $N=1$ (a single spin) then we could reconstruct the
general distribution $P_N(w)$. In fact, let $\hat{P}_N(s)$ denote the
Laplace transform of $P_N(w)$,
\be
\hat{P}_N(s)=\int_0^{\infty} \exp(-Ws)P_N(W)dW
\label{rec1}
\ee
Using the result $W=\sum_{i=1}^Nw_i$ we can write,
\be
\hat{P}_N(s)=\bigl(\hat{P}_1(s)\bigr)^N
\label{rec2}
\ee
allowing us to reconstruct $P_N(w)$ from the sole knowledge of $P_1(w)$.
Although the analytical computation of $P_1(w)$ might be possible by using
other approaches, throughout this paper we have considered a {\em
thermodynamic approach}
where the large-$N$ theory has been taken as an approximation to finite $N$.
This approach turns out to give exact results for all cumulants of the
distribution thereby suggesting that the reconstruction of $P_1(w)$ from
$P_N(w)$ might be possible.  One could naively think that this is
possible just using \eq{rec2} together with the knowledge of
$P_N(w)$. Unfortunately, this is not the case as the knowledge of
$P_N(w)$ is only approximate as we showed in the previous section. There
is however a possible strategy to reconstruct $P_1(w)$ that is based on
the fact that cumulants are exactly known. Let us define the
following function,
\be
h(x)=\lim_{N\to\infty}\frac{g_N(x)}{N}
\label{rec3}
\ee
where $g_N(x)$ was defined in \eq{fse1}. In the large-$N$ limit we can
solve $h(x)$ by applying the saddle-point approximation,
\bea
h(x)=\lim_{N\to\infty}\frac{1}{N}\log\Bigl(\overline{\exp(xW)}\Bigr)=\nonumber\\
\lim_{N\to\infty}\frac{1}{N}\log\Bigl(\int dW \exp(xW)\exp(Ns(w))\Bigr)=xw(x)+s(w(x))~~~.
\label{rec4}
\eea
where $w(x)$ is the solution of the equation,
\be
\frac{ds(w)}{dw}\Bigr|_{w=w(x)}=-x~~~.
\label{rec5}
\ee
For a given value of $w$, \eq{maximum} shows that $x=\lambda(w)$. For
instance, for $x=0,-1/k_BT$ we get $w=w^{\rm mp},w^{\dag}$
respectively. Therefore, we can express $h(x)$ in terms of $w$ rather
than $x$,
\be
h(w)=w\lambda(w)+s(w)
\label{rec6}
\ee
By inserting \eq{fse3} in \eq{rec3} we get $g_1(w)=h(w)$ and therefore,
\be
\int dw'\exp\Bigl(\lambda(w)(w'-w)\Bigr)P_1(w')=\exp\Bigl(s(w)\Bigr)
\label{rec7}
\ee
Formally, this integral equation is closed and provides an
exact solution for $P_1(w)$ in terms of the entropy
$s(w)$. Unfortunately we have been unable to solve it in full generality
(as detailed knowledge of the solution in \eq{rec5} is required). Yet,
for $P_1(w)$ it still holds that there are exponential tails identical
to those we already derived for $P_N(w)$ in the large-$N$ limit. To
show this result we use \eq{maximum} and rewrite \eq{rec7} as follows,
\be
\int dw'\exp\Bigl(-s(w)-\frac{\partial s(w)}{\partial w}(w'-w)\Bigr)P_1(w')=1~~~.
\label{rec8}
\ee
Let us now suppose now that $\lambda(\omega)$ is approximately
constant (equal to $\hat{\lambda}$) showing a plateau over a given
region of work values. From \eq{maximum} then $s(w')\simeq
s(w)+\frac{\partial s(w)}{\partial w}(w'-w)$ and,
\be
P_1(w')\propto \exp(s(w'))=C\exp(\hat{\lambda}w')~~~,
\label{rec9}
\ee
where $C$ is a constant. This shows that the width of the exponential tail
for $P_1(w)$ (and, by extension, for $P_N(w)$ at any value of $N$)
is equal to $\hat{\lambda}$.

\section{The case of magnetic nanoparticles}  
\label{nano}
In this section we discuss a system where the previous theory could be
experimentally tested.  We focus our attention on thermally activated
magnetic nanoparticle systems \cite{review_nano}. These systems have the
great advantage that dynamics is invariant under time-reversal of the
magnetic field $H\to -H$. Also many magnetic field cycles can be
experimentally realized in micro-SQUID machines allowing to
experimentally extract the work distribution with good precision. The
main experimental limitation to observe WF though is the quite large
value of the magnetic moment of the nanoparticle. Transition rates are
described by the Brown-Neel formula,
\be
\tau_{\rm relax}(H)=\tau_0\exp\Bigl( \frac{B(H)}{k_BT} \Bigr) 
\label{a16}
\ee
where $\tau_0$ is a microscopic time describing relaxation within a
state and $B(H)$ is a field dependent barrier.  We consider two cases:
A) paramagnetic molecular clusters where the energy barrier is nearly
field independent $B(H)=B_0$ (this case could also describe specific
ferro and ferrimagnetic nanoparticles where the anisotropy contribution
to the zero-field barrier is negligible, for a discussion see
\cite{GonTej94}); B) ferromagnetic nanoparticles with axial
anisotropy where $B(H)$ depends on the intensity of the external field
projected on the easy magnetization axis as described by the
Stoner-Wohlfarth expression
$B(H)=B_0(1-\Bigl|\frac{H}{H_c}\Bigr|)^{\alpha}$ where $H_c$ is the
field required to suppress the barrier and $\alpha$ is an exponent in
the range $1.5-2$.  Recent experiments have demonstrated how the height
of the barrier $B_0$ can be considerably reduced by applying a
transverse field, making possible to observe
magnetization reversible transitions (also called telegraph noise
measurements) in single Co nanoparticles at low
temperatures~\cite{WerBonHasBenBarDemLoiPasMai97,JamWerThiMaiDupMelPer01}.

As we already discussed in Sec.~\ref{ft}, in a magnetic system a
time-reversal invariant protocol can be accomplished by switching the
magnetic field $H$ from $-H_A$ to $H_A$ ($H_A$ being the amplitude of
the field), the free energy and the rates being an even function of
$H$. Under such conditions work and heat are equivalent if $H_A$
induces a magnetization close to its saturation value. From the
experimental point of view, it is relevant to understand under which
conditions large deviations from the most probable work are
observable. By large deviations we understand work (heat) fluctuations
corresponding to work (heat) values around $w^{\dag}$ ($q^{\dag}$).  A
useful quantity that tell us how difficult it is to sample that region
of work values is the ratio $\Omega$ describing the fraction of
trajectories that transiently violate the second law, $w\le 0$. This
fraction is given by the integrated fluctuation
theorem~\cite{EvaSea02,Crooks98}. This is obtained by rewriting
\eq{ft1},
\be
P_N(-W)=P_N(W)\exp\Bigl(-\frac{W-\Delta
F}{k_BT}\Bigr)=\exp\Bigl(-\frac{W_{\rm dis}}{k_BT}   \Bigr)~~~~.
\label{nano1}
\ee
where we have taken $P_N(W)=P_F(W)=P_R(W)$. Integrating this expression
from $W=0$ up to $W=\infty$ we obtain,
\be
\Omega=\frac{{\cal N}(w<0)}{{\cal N}(w>0)}=\langle\exp(-\frac{Nw}{k_BT})\rangle_{w>0}
\label{a14}
\ee
where ${\cal N}(w<0),{\cal N}(w>0)$ are the fraction of trajectories
for which the total work is negative and positive respectively,
\be
{\cal N}(w<0)=\int_{-\infty}^0dWP(W)~~~~;~~~~{\cal N}(w>0)=\int_0^{\infty}dWP(W)
\label{nano2}
\ee
and the average on the r.h.s of \eq{a14} is restricted to the subset of
trajectories for which $w>0$. Quite generally, we expect that $\Omega$
is a non-universal function dependent on all cumulants of $s(w)$, yet
its exponential dependence in $N$ assures that, in the regime where TV
are observable, $\Omega$ is approximately described by the value of the
average total work divided by the bath temperature $\overline{W}/k_BT$
which is approximately given by $Nw^{\rm mp}/k_BT$.

\begin{figure}
\begin{center}
\epsfig{file=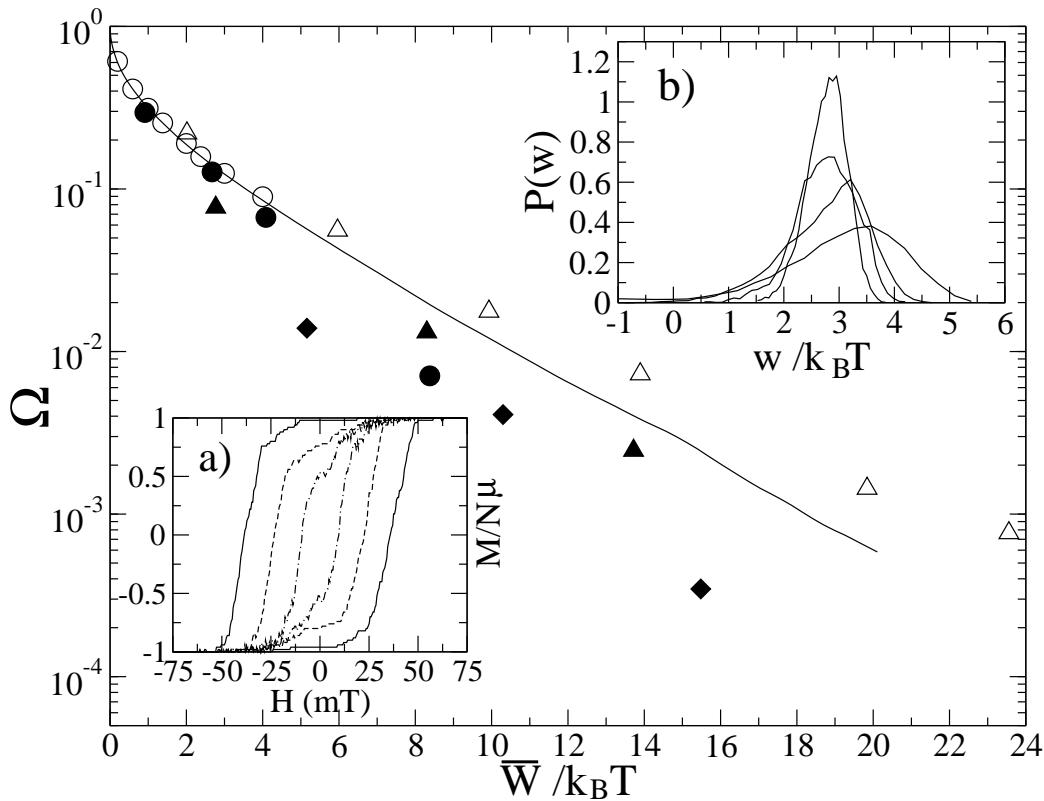,angle=-90, width=14cm}
\end{center}
\caption{\small Application of the theory to magnetic nanoparticles for a case where
$w_{\rm rev}=0$ and $w\simeq q$. Main
panel: $\Omega$ \eq{a14} as a function of $\overline{W}/k_BT$ for Cases
A and B. Number of particles range from $N=1$ up to $20$ for both cases.
Case A) corresponds to nanoparticles (open symbols) with $\mu=300
\mu_B,T=200K,H_A=2{\rm T},\tau_0=10^{-7}s,B_0=2300K$ at ramping speeds
$r=1{\rm mT}/s ({\rm circles}),10{\rm mT}/s ({\rm triangles\, up})$.
Case B) corresponds to ferromagnetic particles (filled symbols) with
$\mu=500\mu_B,T=40K,H_A=238{\rm mT},H_c=119{\rm
mT},B_0=500K,\tau_0=10^{-5}s,\alpha=1.5$ and ramping speeds of $0.01
({\rm circles}),0.1 ({\rm triangles\, up}), 1 ({\rm diamonds}) {\rm
T}/s$ (continuous,dashed and dotted lines respectively). The continuous
line is the prediction for a Gaussian work distribution (see discussion
at the end of Sec.~\ref{thermo}) in the
linear-response regime $\s_w^2=2k_BT\overline{w_{\rm dis}}$ ($R=1$ in
\eq{mp5}). Insets: Both are for ferromagnetic nanoparticles (Case B)
with the same parameters as in the main panel. Inset a) shows hysteresis
cycles for $N=100$ ferromagnetic nanoparticles at the three ramping
speeds (larger hysteresis for higher speeds). Inset b) shows dissipated
work distributions at ramping speeds $0.1 {\rm T}/s$ for $N=1,5,10,20$
particles (larger sizes correspond to narrower distributions).}
\label{fig12}
\end{figure}

We choose Glauber rates as these have been experimentally demonstrated
to describe very well the relaxation of single magnetic
moments~\cite{Can01,Cou04}. These are given by \eq{num1} where $\tau_{\rm
relax}(H)$ is given by \eq{a16}.  We consider ramping
experiments~\cite{RAMPING} where $N$ particles are subject to the
action of a field that is switched from $H=-H_A$ up to $H=H_A$ at a
constant speed $\dot{H}$. We generate individual trajectories
according to the Glauber rates by starting from initial configurations
with $M=M_{\rm eq}(H_A)$ and repeating the ramping protocol many
times, each time the total work \eq{a2} is computed,
$W=-\mu\int_{-H_A}^{H_A}M(H)dH$.  If $H_{\rm sw}$ is the field at
which the magnetization of a given particle switches for the first
time then, for a given trajectory, some of the particles will switch
state at a value of the field $H_{\rm sw}<0$, while others will switch
at $H_{\rm sw}>0$. For fast ramping speeds the dynamics is well
described by a first-order Markov process~\cite{HanTalBor90} and the
dissipated work for that trajectory will be identical to the value
$2H_{\rm sw}$ averaged over all particles. In general, for lower
ramping speeds, the relation between the dissipated work and the value
of $H_{\rm sw}$ is more complicated. To estimate $\Omega$ we generate
trajectories and evaluate the fraction of them with $W>0$ and
$W<0$. We chose to do numerical simulations rather than applying the
large-$N$ theory to give a more clear picture about which results can
we expect from a finite number of ramping experiments (around 10000).
In the main panel of Figure~\ref{fig12} we plot the value of $\Omega$
\eq{a14} as obtained for different ramping protocols in cases A and
B. All points scatter around a generic (but non-universal) curve
useful to predict in which regime TV are expected to be
observable. An important advantage of the time-reversal symmetry
property $H\to -H$ of magnetic nanoparticle systems, as compared to
other systems~\cite{LipDumSmiTinBus02,WanSevMitSeaEva02}, is the
feasibility of performing many ramping cycles in a single experiment
making TV observable for $\Omega$ values as low as
$10^{-4}$. According to Figure~\ref{fig12}, TV should be observable
for work values as large as $20k_BT$.

\section{Conclusions.} 
\label{conclusions}
Two-state systems provide a simple conceptual framework to analyze
work fluctuations (WF) and transient violations (TV) of the second
law.  These non-equilibrium effects are expected to be relevant and
observable for nanosized objects when the energies involved are
several times $k_BT$, $k_B$ being the Boltzmann constant and $T$ the
temperature of the bath. These have been already observed in the
unfolding of small RNA hairpins~\cite{LipDumSmiTinBus02} as well as in
polysterene beads dragged through a solvent
\cite{WanSevMitSeaEva02}. Related measurements include the
experimental test of the Gallavotti-Cohen fluctuation theorem in
Rayleigh-Bernard convection \cite{CilLar98} and turbulent flows
\cite{CilGarHerLacPinRui03}.  Other experiments include the
observation of gravitatory potential energy fluctuations in driven
granular media~\cite{FeiMen04}.  The scientific discipline behind all
such rich phenomenology deserves to be called {\em thermodynamics of
small systems}.  It deals with the thermal behavior of non-equilibrium
small systems where the typical energies are few times $k_BT$. The
statistics of energy exchange processes between the system and the
thermal environment is described by frequent Gaussian distributed
events plus rare events corresponding to large statistical deviations
from the average value. The theoretical and experimental study of
these fluctuations could be of relevance to understand issues related
to the organization and function of biological matter in the
nanoscale~\cite{Ritort_paris}.

In this paper we studied WF in two-state systems. We have introduced a
trajectory thermodynamics formalism with the specific aim to quantify WF
in such model.  We have shown how to define a trajectory entropy $s(w)$
that characterizes WF around the most probable value $w^{\rm mp}$, and a
trajectory free-energy ${\cal F}(w)$ whose minimum value at $w=w^{\dag}$
specifies the value of the work that needs to be efficiently sampled to
quantitatively test the Jarzynski equality.  The theory requires the
introduction of a Lagrange multiplier $\lambda(w)$, its inverse
playing the role of a temperature in the trajectory thermodynamics
formalism. Analytical expressions for the trajectory potentials
$s(w),{\cal F}(w)$ have been also derived.  In general, both values $w^{\rm
mp}$ and $w^{\dag}$ are of the same magnitude but opposite sign, meaning
that large deviations of WF need to be sampled to recover equilibrium
free-energies from non-equilibrium measurements, e.g. by using the
Jarzynski equality.

We have then carried out a systematic study of WF in the framework of
the large-$N$ theory.  Several results are worth mentioning. First of
all, we have found an analytical expression for the trajectory entropy
that satisfies the fluctuation theorem by Crooks \cite{Crooks98} that
relates forward and reverse processes. An important result is that the
value of the work $w^{\dag}$ that has to be sampled in order to test
the Jarzynski equality is equal to the most probable value of the work
(with a minus sign) for the reverse process. Intuitively this means
that the forward and reverse distributions must overlap each other in
order to get good estimates of the work using the Jarzynski equality,
a result that was emphasized long-time ago by
Bennett~\cite{Bennett76}. Furthermore, if both forward and reverse
processes are symmetric mirror images then $w^{\rm rev}=0$ and
$w^{\dag}=-w^{\rm mp}$ independently of how far the system is driven
out of equilibrium . This last case is particularly interesting as the
total work practically coincides with the heat. The fluctuation
theorem by Crooks is then also applicable to the heat in that limit, a
result that is quite reminiscent of a heat fluctuation theorem
recently derived~\cite{ZohCoh03,ZonCilCoh03}. For the heat
distribution, we find that it is described by a central Gaussian
distribution describing {\em local equilibrium}, i.e. with $R=1$, and
long exponential tails with widths described by the Lagrange
multiplier $\lambda(w)$, which plays the role of the inverse of a
temperature. Strictly speaking, because the temperature must be a
positive quantity, only the tails in the negative sector $q<<-1$ where
$\lambda$ is negative admit such an interpretation (i.e. in the sector
of WF dominated by TV). It has not escaped our attention that this
temperature could be related to other non-equilibrium temperatures
that have been defined in other contexts
~\cite{CasJou03,CriRit03}, such as steady-state~\cite{Zam04} or aging
systems~\cite{CriRit04}.

Our study raises the following question: to what extent are work and
heat fluctuations equivalent? We already emphasized in Sec.~\ref{heat}
that work and heat should be equivalent, at least this is the
underlying content of the first law of thermodynamics. However, from
the perspective provided by the present analysis, some important
differences can be underlined. Exponential tails are more often
observed in the heat rather than in the work. Such result has been explicitly
shown in the case of a bead dragged through a fluid \cite{ZohCoh03}
where the work is clearly Gaussian distributed while the heat displays
exponential tails. However, in that case the origin of this difference
lies on the fact that the motion for the bead is described by a
stochastic linear equation which in general might not be the case. The
difference between heat and work has its root at the true microscopic
definition of these quantities. Heat is identical to work when the final
energy of the system is constrained be identical to the initial value
(i.e. $Q=W$ if $\Delta E=0$, for the heat we adopt the sign convention
of Sec.~\ref{heat}). The simplest interpretation is that exponential
tails in the work distribution are always present if the model is
non-linear by definition (which is not the case for the aforementioned
case of the bead dragged through the fluid). However, work
distributions always tend to be masked by a Gaussian contribution
coming out from the Gaussian fluctuations that characterize the
initial equilibrium state.  Therefore, only when thermal fluctuations
in the initial and final states are negligible as compared to the
total amount of work along the trajectory, the measured work
distributions are paralleled by the heat distributions and tails can
be observed. This explains the qualitative difference observed
between the functions $\lambda(w)$ in Figs.~\ref{fig6ab2} and the
right panel in Fig.~\ref{fig3}. In the latter, Gaussian fluctuations
in the energy of the initial and final configurations tend to mask the
presence of the exponential tails.

We also studied finite-size effects to test how good the large-$N$
theory is and provided a strategy to re-derive the finite-$N$ work
distribution from the large-$N$ result. An important conclusion is
that the large-$N$ theory accounts for the existence of exponential
tails also at finite $N$, the value of the widths
$\lambda_+,\lambda_-$ (corresponding to the plateaus in $\lambda(w)$)
being independent of $N$.  In addition, we applied the theory to
magnetic nanoparticle systems which provide an experimental
realization of two-state systems. We studied under which conditions
the theory can be experimentally tested. Our results suggest that WF
and TV should be observable whenever average work values are not much
larger than $20k_BT$. It is realistic to say that we are currently at
the limit of the resolution of current micro-SQUID devices for the
detection of single small magnetic moments (around few hundreds of
$\mu_B$). Surely, we will see developments in the near future and
experimental measurements of WF in magnetic systems, as well as the
test of the present theory, will become possible.

{\bf Acknowledgments.} The author acknowledges the warm hospitality of
Bustamante and Tinoco labs at UC Berkeley where this work was done. He
is grateful to C. Bustamante, J. Gore, C. Jarzynski, A. Labarta and I. N. Tinoco for
useful discussions. This work has been supported by the David and Lucile
Packard Foundation, the Spanish Ministerio de Ciencia y Tecnolog\'{\i}a
Grant BFM2001-3525 and Generalitat de Catalunya.


\begin{thebibliography}{99}

\bibitem{EvaCohMor93} Evans, D., J., Cohen, E., G., D., \& Morriss, G., P. (1993) {\em Phys. Rev. Lett.} {\bf 71}, 2401-2404 

\bibitem{EvaSea02} Subsequent work has been reviewed in 
Evans, D. \& Searles, D. (2002) {\em Adv. Phys.} {\bf 51}, 1529

\bibitem{Jarzynski97} Jarzynski, C. (1997) {\em Phys. Rev. Lett.} {\bf
78}, 2690

\bibitem{Kurchan98} Kurchan, J. (1998) {\em J. Phys. A (Math. Gen.)} {\bf 31}, 3719 

\bibitem{Crooks98} Crooks, G. E. (1998) {\em J. Stat. Phys.} {\bf 90}, 1481; (2000) {\em Phys. Rev. E} {\bf 61}, 2361  

\bibitem{HumSza01} Hummer, G., \& Szabo, A. (2001) {\em Proc. Natl. Acad. Sci. USA} {\bf 98} 3658-3661




\bibitem{LipDumSmiTinBus02} Liphardt, J., Dumont, S., Smith, S., B., Tinoco,
I., Jr., \& Bustamante, C. (2002) {\em Science} {\bf 296}, 1832-1835

\bibitem{WanSevMitSeaEva02} Wang, G., M., Sevick, E., M., Mittag, E., Searles, D., J., \& 
Evans, D. , J. (2002) {\em Phys. Rev. Lett.} {\bf 89}, 050601


\bibitem{CilLar98} Ciliberto, S. \& Laroche, C. (1998) {\em J. Phys. IV
(France)} {\bf 8} 215

\bibitem{CilGarHerLacPinRui03} Ciliberto, S., Garnier, N., Hernandez,
S., Lacpatia C., Pinton J.-F. \& Ruiz Chavarria G. (2003) {\em Preprint
arXiv:nlin.CD/0311037v2}


\bibitem{LauPinSchStoWol00} Laughlin, R., B., Pines, D., Schmalian, J.,
Stojkovic, B., P., \& Wolynes, P. (2000) {\em Proc. Nat. Acad. Sci. USA} {\bf 97}, 32

\bibitem{RitBusTin02} Ritort, F., Bustamante, C., \& Tinoco, I., Jr. (2002)
{\em Proc. Nat. Acad. Sci. USA} {\bf 99}, 13544 

\bibitem{GalCoh95} Gallavotti, G., Cohen, E., G., D., (1995) {\it J. Stat. Phys.} {\bf 80}, 931


\bibitem{ZucWoo02} Zuckerman, D., M., \& Woolf, T., B. (2002)
{\em Chem. Phys. Lett.} {\bf 351}, 445; (2002) {\em Phys. Rev. Lett.} {\bf 89}, 180602

\bibitem{GorRitBus03} Gore, J., Ritort, F., \& Bustamante, C.
(2003) {\em Proc. Nat. Acad. Sci. USA} {\bf 100}, 12564

\bibitem{MazJar99} Mazonka. O. \&  Jarzynski, C. (1999) {\em
Preprint arXiv:cond-mat/9912121}

\bibitem{ZohCoh03} Van Zon, R., \& Cohen, E., G., D. (2003) {\em Phys. Rev. Lett.} {\bf
91}, 110601 ;  (2003) {\em Phys. Rev. E} {\bf 67}, 046102 

\bibitem{CriRit04} Crisanti, A. \& Ritort, F. (2004) {\em Europhys. Lett.} {\bf 66}, 253

\bibitem{Ritort03} Ritort, F. {\em Preprint arXiv:cond-mat/0311370}
Proceedings of the workshop ``Unifying concepts in granular media and
glasses''. Eds. A. Coniglio, A. Fierro, H. J. Hermann and M. Nicodemi,
Elsevier-Amsterdam (2004).

\bibitem{Ritort04} Ritort, F. (2004) {\em J. Phys. Chem. B} {\bf 108}, 6893


\bibitem{BelCilLar01} Buisson, L., Bellon, L., \& Ciliberto, S. (2003)
{\em J. Phys. C (Cond. Matt.)} {\bf 15}, S1163

\bibitem{Cip03} L. Cipelletti et al., J. Phys. C ({\em Cond. Matt.}) {\bf 15}, S257 (2003)  

\bibitem{review_nano} Wernsdorfer, W. (2001) {\em Adv. Chem. Phys.} {\bf 118}, 99 


\bibitem{GonTej94} Gonzalez-Miranda, J., M., \&  Tejada, J. (1994) {\em Phys. Rev. B} {\bf 49}, 3867 

\bibitem{WerBonHasBenBarDemLoiPasMai97} Wernsdorfer, W., Bonet-Orozco, E., Hasselbach, K., Benoit, A., Barbara, B., Demoncy, N., Loiseau, A., Pascard, H. \& Mailly, D. (1997) {\em Phys. Rev. Lett.} {\bf 78}, 1791

\bibitem{JamWerThiMaiDupMelPer01} Jamet, M., Wernsdorfer, W., Thirion, C., Dupuis, V., M\'elinon, P., \& P\'erez, A. (2001) {\em Phys. Rev. Lett.} {\bf
86}, 4676 

\bibitem{Can01} Caneschi, A. et al. (2001) {\em Preprint arXiv:cond-mat/0106224}


\bibitem{Cou04} Coulon, C. et al. (2004) {\em Preprint arXiv:cond-mat/0404620}


\bibitem{RAMPING} Kurk\"{\i}jarvi, J. (1972) Phys. Rev. B {\bf 6}, 832; Gunther, L., \& Barbara, B. (1994) {\em Phys. Rev. B} {\bf 49}, 3926; Garg, A. (1995) {\em Phys. Rev. B} {\bf 51}, 15592

\bibitem{HanTalBor90} Hanggi, P., Talkner, P., \& Borkovec, M. (1990)
{\em Rev. Mod. Phys.} {\bf 62}, 251

\bibitem{FeiMen04} Feitosa, K. \& Menon, N. (2004) {\em
Phys. Rev. Lett.} {\bf 92}, 164301 

\bibitem{Ritort_paris} Ritort, F. (2003) {\em Seminaire Poincar\'e}, {\bf
2}, 63; {\em Preprint arXiv:cond-mat/0401311} 

\bibitem{Bennett76} Bennett, C., H., (1976) {\em J. Comp. Phys.} {\bf 22}, 245

\bibitem{ZonCilCoh03} Van Zon, R., Ciliberto, S., Cohen, E., G.,
D. (2003) {\em Preprint arXiv: cond-mat/0311629}

\bibitem{CasJou03}  For a review, Casas-Vazquez, J., \& Jou, D. (2003) {\em
Rep. Prog. Phys.} {\bf 66}, 1937


\bibitem{CriRit03} For a review, Crisanti, A. \& Ritort, F. (2003) {\em J. Phys. A
(Math. Gen.)} {\bf 36}, R181 

\bibitem{Zam04} Zamponi, F., Ruocco, G., \& Angelani, L. (2004) {\em
Preprint arXiv: cond-mat/0403579}

\end{thebibliography}
\end{document}